\documentclass[superscriptaddress,amsmath,amssymb,prd,preprintnumbers,showpacs,twocolumn,longbibliography]{revtex4-2}

\usepackage{dutchcal}
\usepackage{hyperref}
\usepackage{amssymb,amsmath}
\usepackage{url}
\usepackage{dsfont}
\usepackage{units}
\usepackage{booktabs}
\usepackage[dvipsnames,table,xcdraw]{xcolor}
\usepackage{graphicx}
\usepackage{epsfig}
\usepackage{float}
\usepackage{mathrsfs}
\usepackage{slashed}
\usepackage{cancel}
\usepackage{units}
\usepackage{upgreek}
\usepackage{subfig}

\newcommand{\beq}{\begin{equation}}
\newcommand{\eeq}{\end{equation}}

\newcommand\vex[1]{\mathbf{#1}}

\def\ket#1{\mathinner{|{#1}\rangle}}

\usepackage[T3,T1]{fontenc}
\DeclareSymbolFont{tipa}{T3}{cmr}{m}{n}
\DeclareMathAccent{\invbreve}{\mathalpha}{tipa}{16}

\newlength{\hhatheight}

\def\ring#1{{\mathaccent'27 #1}}

\usepackage[defaultcolor=blue]{changes}
\definechangesauthor[name=Babak, color=orange]{b}

\definechangesauthor[name=Ralf, color=purple]{r}

\usepackage{xcolor}

\makeatother

\graphicspath{{./figures/pdf/}}

\newcommand{\orcid}[1]{\href{https://orcid.org/#1}{\includegraphics[width=10pt]{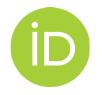}}}

\begin{document}

\title{Lorentz violation and momentum-space geometric phases}

\author{V.\ Alan Kosteleck\'y\orcid{0000-0003-1610-8094}}
\affiliation{Department of Physics, Indiana University, Bloomington, Indiana 47405, USA}
\affiliation{Indiana University Center for Spacetime Symmetries, Bloomington, Indiana 47405, USA}

\author{Ralf Lehnert\orcid{0000-0001-6682-5409}}
\affiliation{Department of Physics, Indiana University, Bloomington, Indiana 47405, USA}
\affiliation{Indiana University Center for Spacetime Symmetries, Bloomington, Indiana 47405, USA}

\author{Marco Schreck\orcid{0000-0001-6585-4144}}
\affiliation{Departamento de F\'{i}sica, Universidade Federal do Maranh\~{a}o, Campus Universit\'{a}rio do Bacanga, S\~{a}o Lu\'{i}s (MA), 65085-580, Brazil}

\author{Babak Seradjeh\orcid{0000-0003-1834-8286}}
\affiliation{Department of Physics, Indiana University, Bloomington, Indiana 47405, USA}
\affiliation{Indiana University Center for Spacetime Symmetries, Bloomington, Indiana 47405, USA}
\affiliation{Quantum Science and Engineering Center, Indiana University, Bloomington, Indiana 47405, USA}

\begin{abstract}

Geometric phases 
can manifest when a relativistic quantum particle
moves cyclically along a loop in parameter space.
The phase can be affected by the presence of a background field
and can be accompanied by nontrivial topological features.
The appearance of adiabatic geometric phases in momentum space
is demonstrated for a Lorentz-violating Weyl fermion,
where the role of the background is played
by the coefficients for Lorentz violation.
As explicit examples,
the Berry curvature and the first Chern number are derived
for two cases with large Lorentz violation,
one incorporating CPT violation and one preserving CPT symmetry.
Some alternative topological invariants are also obtained.
In certain scenarios with large Lorentz violation,
the physical vacuum is associated with a topological phase.

\end{abstract}


\maketitle

\section{Introduction}

As a quantum state evolves, 
it develops a dynamical phase governed by a hamiltonian.
The state can also acquire an additional observable geometric phase 
when parameters of the system undergo an adiabatic change 
along a closed curve in parameter space.
This additional phase,
known as the Pancharatnam-Berry phase 
or Berry phase~\cite{Pancharatnam:1956url,Berry:1984jv},
has an interpretation as the holonomy of the curve 
in the U(1) bundle over the parameter space~\cite{Simon:1983mh},
and in a nontrivial bundle
the phase is associated with topological invariants. 
Geometric phases are known to appear in other scenarios,
such as nonadiabatic motion~\cite{aa87}
or modified boundary conditions~\cite{jz89},
and the appearance of geometric and topological phases in physics is now
widespread~\cite{Chruscinski:2004,Nakahara:2003nw,Vanderbilt:2018}. 

The parameter space associated with the Berry phase
can be spanned by quantities such as applied fields 
that appear in the hamiltonian of the system.
Alternatively, 
the parameter space can be taken as three-dimensional momentum space
$\mathcal{P}_3$,
and the wave function at position $\mathbf{r}$ written as 
$\exp(\mathrm{i}\vex{p}\cdot{\vex{r}})\ket{\psi(\vex{p})}$ 
for $\mathbf{p}\in\mathcal{P}_3$.
The quantum state $\ket{\psi(\vex{p})}$
resides in the U(1) bundle over $\mathcal{P}_3$,
and it measures the deviation of the wave function from a plane wave 
as determined from the structure of the hamiltonian
and the boundary conditions. 
Adiabatic evolution of the wave function 
around a closed curve in $\mathcal{P}_3$
can then generate a nontrivial momentum-space geometric phase. 

The focus of this work is the effect on the momentum-space geometric phase 
of a relativistic fermion moving in a generic background.
This physical scenario is central 
to theoretical and experimental studies of Lorentz violation.
In the context of fundamental physics,
for example,
Lorentz violation represents a prospective signal 
for an underlying unified theory at the Planck scale
such as strings~\cite{Kostelecky:1988zi,Kostelecky:1991ak}. 
Tiny corrections to known physics can be incorporated 
using effective field theory~\cite{Weinberg:2009bg},
yielding a framework known as the Standard-Model Extension (SME)
that describes general effects 
of Lorentz violation in particle physics and 
gravity~\cite{Kostelecky:1994rn,Colladay:1996iz,Colladay:1998fq,%
Kostelecky:2003fs,Kostelecky:2020hbb}.
As another example,
in condensed-matter physics
the treatment of Weyl and Dirac semimetals is closely analogous to
the effective field theory of Lorentz-violating fermions.
The dynamical behavior of a quasiparticle excitation
in certain regions of the semimetal band structure
displays measurable deviations 
from emergent Lorentz invariance~\cite{Lv:2015pya,PhysRevX.6.031021,%
Yan:2016euz,Armitage:2017cjs,gvkr19},
and Lorentz-violating effective field theory 
can be used to describe the associated
physics~\cite{Kostelecky:2021bsb,Gomez:2023jyl,%
AlanKostelecky:2024gek,AlanKostelecky:2024psy}.

In the SME,
each Lorentz-violating term in the Lagrange density
is accompanied by a coefficient
that acts as a background controlling observable effects.
The coefficient can appear in the quantum hamiltonian 
and hence can modify the geometric phase of the fermion
when it moves adiabatically along a curve in momentum space,
even when the coefficient itself is homogeneous and constant in time.
The general effective field theory for a Lorentz-violating fermion
thus represents an ideal framework for investigating
the background dependence of momentum-space geometric phases.
Here, 
we explore this background dependence
by investigating two examples
that are widely used in studies of Lorentz-violating physics.
One has a background involving 
a coefficient $b_\mu$ for Lorentz violation
that incorporates also violation of the product CPT
of charge conjugation C, parity inversion P, and time reversal T.
The background in the second example involves 
a coefficient $d_{\mu\nu}$ for Lorentz violation
that preserves CPT symmetry.

The relationship between the background 
and the momentum-space geometric phase is of interest in its own right.
However,
it can have particular relevance for scenarios with large Lorentz violation, 
where perturbative techniques are inadequate to understand the physics.
These scenarios arise
either when the coefficients for Lorentz violation
are intrinsically large in a chosen frame
or when they are small in one frame but a highly boosted frame is adopted.
Identifying a consistent framework for treating these scenarios
is known as the concordance problem~\cite{Kostelecky:2000mm}.
For a fermion in the presence of Lorentz violation
governed by dominant operators of mass dimension three,
a generic physical solution to the concordance problem
has recently been demonstrated using thermodynamic arguments
and the properties of semimetals~\cite{AlanKostelecky:2024gek}.
In this picture,
the physical vacuum at zero temperature
contains particles with negative energies,
paralleling the situation in the ground state of certain Weyl
semimetals~\cite{Grushin:2012mt,Behrends:2018qkj,Kostelecky:2021bsb}.
These particles endow the physical vacuum 
with a momentum-space geometric phase
that has topological properties determined by the background.

The momentum-space geometric phases studied in this work
involve relativistic dispersions
for time-independent and spatially homogeneous backgrounds,
including ones with large Lorentz violation
and a nontrivial physical vacuum.
Other types of geometric phases associated with Lorentz violation
can also be considered.
For example,
in the limit of nonrelativistic motion and perturbative Lorentz violation,
the spatial components $b_j$ of the coefficient $b_\mu$
couple to the fermion spin like a background magnetic 
field~\cite{Kostelecky:1999zh},
so the canonical illustration of the Berry phase
arising when a nonrelativistic spin evolves
in an adiabatically varying magnetic field~\cite{Berry:1984jv}
therefore maps directly to this scenario.
The case with a perturbative coefficient $d_{\mu\nu}$ 
has been studied by Casana {\it et al.}~\cite{Casana:2015kna},
who show that the corresponding nonrelativistic hamiltonian
induces an Aharonov-Anandan phase~\cite{aa87}
in the wave function of an electron moving on a one-dimensional ring.
Examples of geometric phases for nonrelativistic fermion motion
in the presence of nonminimal SME operators of dimensions five or more
coupled to the electromagnetic field have been extensively studied
as well~\cite{Belich:2004ng,Belich:2006tk,Ribeiro:2007fe,%
Bakke:2011zz,Belich:2011dz,Bakke:2012gt,Bakke:2012qj,%
Bakke:2013lqa,deLima:2013rhe,Bakke:2014gfa, Bakke:2024peu}.

The paper is organized as follows. 
Basic results are outlined for the Lorentz-violating fermions
with coefficients $b_\mu$ and $d_{\mu\nu}$
in Sec.~\ref{lvbasics}
and for momentum-space geometric phases
in Sec.~\ref{msgphase}.
The derivations of the the momentum-space geometric phases
and the associated Berry curvatures and first Chern numbers
in the presence of $b_{\mu}$ are discussed 
in Sec.~\ref{sec:application-sme-b},
while the analogous results are obtained for $d_{\mu\nu}$
in Sec.~\ref{sec:application-sme-d}. 
Some alternative topological invariants
are considered in Sec.~\ref{sec:alternative-invariants}.
We conclude in Sec.~\ref{sec:conclusions}. 
Our conventions incorporate
natural units with $\hbar=c=1$ unless otherwise stated.
Greek letters are adopted for Lorentz indices,
and Latin letters represent spatial indices.
The Minkowski metric $\eta_{\mu\nu}$ has negative signature,
and the Dirac matrices $\gamma^{\mu}$ satisfy the Clifford algebra 
$\{\gamma^{\mu},\gamma^{\nu}\}=2\eta^{\mu\nu}$,
with $\gamma_5:=\mathrm{i}\gamma^0\gamma^1\gamma^2\gamma^3$.

\section{Basics}

\subsection{Lorentz-violating fermions}
\label{lvbasics}

The full SME incorporates general background couplings
to all fields appearing in the Standard Model and General 
Relativity~\cite{Kostelecky:1994rn,Colladay:1996iz,Colladay:1998fq,%
Kostelecky:2003fs,Kostelecky:2020hbb}.
Numerous experimental studies of these backgrounds
have been performed in recent years,
in some cases
achieving Planck-scale sensitivities or beyond
to the coefficients for Lorentz 
violation~\cite{Kostelecky:2008ts}.
Reviews of the SME can be found, for example, in
Refs.~\cite{Bluhm:2005uj,Tasson:2014dfa,Will:2014kxa,%
Hees:2016lyw,Roberts:2021vsi,Bailey:2023pfd}.
For the purposes of the present work,
we limit attention to a Lorentz-violating fermion 
moving in flat spacetime in the presence of
backgrounds that are time independent and spatially homogeneous.

Our focus here is on two specific scenarios, 
one involving a coefficient $b_\mu$ 
and the other a coefficient $d_{\mu\nu}$.
Each coefficient modifies the hyperboloidal mass shell
for a massive Dirac fermion or the standard light cone for a Weyl fermion.
In both cases the modifications are spin dependent,
so the spinor degrees of freedom in the energy eigenstates
can play a significant role in the analysis.
For $b_\mu$ the effects include CPT violation,
but for $d_{\mu\nu}$ they are CPT invariant.

The action for a Dirac spinor field $\psi$ of mass $m_{\psi}$
coupled to the coefficient $b_{\mu}$ for Lorentz violation
can be written as~\cite{Colladay:1996iz,Colladay:1998fq,Kostelecky:2000mm}
\begin{equation}
S_b=\int\mathrm{d}^4x\,\tfrac{1}{2}
\overline{\psi}(\gamma^{\mu}\mathrm{i}\partial_{\mu}-m_{\psi}
-b_{\mu}\gamma_5\gamma^{\mu})\psi+\text{h.c.}\,,
\label{bact}
\end{equation}
where $\overline{\psi}:=\psi^{\dagger}\gamma^0$ as usual.
Note that $b_\mu$ has mass dimension one in natural units.
The properties of this action have been explored 
in the context of quantum field theory in fundamental 
physics~\cite{Colladay:1996iz,Jackiw:1999yp,Perez-Victoria:1999erb,%
Chung:1998jv,Chung:1999pt,Chung:1999gg,Kostelecky:2001jc,%
Altschul:2003ce,Altschul:2004gs,Lehnert:2004ri,Altschul:2004wq,%
Altschul:2005mu,Lehnert:2006id,BaetaScarpelli:2008jsv,%
Kostelecky:2013rta,BaetaScarpelli:2013rmt,Assuncao:2016fko,%
Reis:2016hzu,Kostelecky:2018yfa,Ferrari:2018tps,%
Goncalves:2018vlk,Gomez:2022yrp,AlanKostelecky:2024psy} 
and in the contexts of classical mechanics 
and Finsler geometry~\cite{Kostelecky:2010hs,Kostelecky:2011qz,%
AlanKostelecky:2012yjr,Javaloyes:2013ika,Colladay:2015wra,Foster:2015yta,%
Colladay:2017bon,Silva:2019qzl,Davis:2025}.
The action $S_b$ has also been widely studied
as a phenomenological theory for Lorentz violation
in electrons, protons, neutrons, and other fermions
in atomic, nuclear, and particle physics and in astrophysics%
~\cite{Bluhm:1998rk,Bluhm:1999ev,Bluhm:1999dx,%
Kostelecky:1999mr,Mittleman:1999it,Dehmelt:1999jh,Bear:2000cd,%
Phillips:2000dr,Humphrey:2001wm,Muong-2:2001xzf,Hughes:2001yk,Hou:2003zz,%
Heckel:2006ww,Muong-2:2007ofc,Heckel:2008hw,Altarev:2009wd,%
Brown:2010dt,Gemmel:2010ft,Peck:2012pt,Fittante:2012ua,%
Allmendinger:2013eya,Roberts:2014dda,Stadnik:2014xja,%
Gomes:2014kaa,Noordmans:2014hxa,Roberts:2014cga,Berger:2015yha,%
Kostelecky:2015nma,Ding:2016lwt,%
BASE:2016yuo,BASE:2017rlv,Escobar:2018hyo,Smorra:2019qfx,Ding:2019aox,%
Smorra:2020nko,Ding:2020aew,Acevedo-Lopez:2023wkv,Belyaev:2024chj},
while in condensed-matter physics it describes
the band structures of certain Weyl semimetals%
~\cite{Kostelecky:2021bsb,Gomez:2023jyl,Grushin:2012mt,Zyuzin:2012vn,%
Landsteiner:2015pdh,kk16,rj16,kw17,jw17,Behrends:2018qkj,Lv:2021oam}.

The dimensionless coefficient $d_{\mu\nu}$ 
governs a Lorentz-violating operator
similar to that for $b_\mu$ but containing an additional derivative.
The corresponding action takes the 
form~\cite{Colladay:1996iz,Colladay:1998fq,Kostelecky:2000mm}
\begin{equation}
S_d=\int\mathrm{d}^4x\,\tfrac{1}{2}
\overline{\psi}(\gamma^{\mu}\mathrm{i}\partial_{\mu}-m_{\psi}
+d_{\mu\nu}\gamma_5\gamma^{\mu}\mathrm{i}\partial^{\nu})\psi+\text{h.c.}\,.
\label{dact}
\end{equation}
Features of this action have been investigated in quantum field theory%
~\cite{Kostelecky:2001jc,Altschul:2004wq,Colladay:2012rv,%
Kostelecky:2013rta,Russell:2015gwa,Reis:2016hzu,Reis:2017ayl,%
AndradedeSimoesdosReis:2019aim,Reis:2021ban},
and constraints on the components of $d_{\mu\nu}$ 
have been obtained from experiments in several subfields%
~\cite{Kostelecky:1999mr,Muong-2:2001xzf,Cane:2003wp,%
Altschul:2006he,Muong-2:2007ofc,Heckel:2008hw,Altschul:2009iz,%
Fittante:2012ua,Fittante:2012ua,D0:2012rbu,Stadnik:2014xja,%
Roberts:2014cga,Schreck:2017isa,Lunghi:2020hxn,%
Acevedo-Lopez:2023wkv,Cabral:2024ykw,Aghababaei:2024hcb}.
Note that $d_{\mu\nu}$ has 16 components.
However,
the trace $\eta^{\mu\nu}d_{\mu\nu}$ is an observer scalar
controlling only Lorentz-invariant effects,
so in what follows we limit attention 
to the 15 nontrace components of $d_{\mu\nu}$.

In the action $S_b$,
values of $b_\mu$ large compared to the fermion mass $m_\psi$
represent large Lorentz violation 
and can be associated with qualitative changes to the physical
vacuum~\cite{AlanKostelecky:2024gek}.
For completeness,
it is thus desirable to avoid the assumption of perturbative Lorentz violation
when exploring the momentum-space geometric phases.
We therefore choose to work here in the massless limit $m_\psi\to0$,
which ensures that the Lorentz violation is always large 
in the action $S_b$
and simplifies some of the analysis for the action $S_d$.
This limit retains physically meaningful Lorentz violation
provided the fermion $\psi$ is implicitly understood 
to have suitable interactions obstructing any field redefinitions
that could otherwise eliminate 
the coefficients for Lorentz violation~\cite{Colladay:1998fq}.
These interactions leave unaffected 
our analysis of momentum-space geometric phases,
which is based on the motion of particle states 
described by the quadratic actions $S_b$ and $S_d$.

\subsection{Momentum-space geometric phases}
\label{msgphase}

The momentum-space geometric phase of interest here
arises when a quantum state vector $\ket{\psi(\vex{p})}$ in the Hilbert space 
undergoes adiabatic motion along a curve $\mathcal{C}$ 
in three-dimensional momentum space $\mathcal{P}_3$.
In geometric terms,
this motion can be understood as parallel transport
in the U(1) complex vector bundle of states over $\mathcal{P}_3$.
The parallel transport is described by the Berry connection 1-form, 
which in this case has three components given by 
\begin{equation}
\mathcal{A}_{\psi}(\mathbf{p})
:= \mathrm{i}\langle \psi(\mathbf{p})|\nabla_{\mathbf{p}}
|\psi(\mathbf{p})\rangle\,.
\label{eq:berry-connection}
\end{equation}
This connection is related to the position operator.
In a crystalline solid,
for example,
$\vex p$ can be taken as the lattice momentum
with $\ket{\psi(\vex p)}$ as the Bloch state,
and the action of the position operator
in the momentum space spanned by 
$\exp(\mathrm{i}\vex{p}\cdot{\vex{r}})\ket{\psi(\vex{p})}$ 
then matches 
$\mathrm{i}\nabla_{\mathbf{p}} + \mathcal{A}_{\psi}(\mathbf{p})$.

The Berry connection is unphysical
because it changes under a gauge transformation of the wave function,
\begin{subequations}
\begin{align}
\psi(\mathbf{p})\rangle &\mapsto 
|\psi'(\mathbf{p})\rangle
=\exp(\mathrm{i}\alpha(\mathbf{p}))|\psi(\mathbf{p})\rangle\,,
\\[2pt]
\mathcal{A}_\psi &\mapsto 
\mathcal{A}_{\psi'} 
= \mathcal{A}_\psi - \nabla_{\mathbf{p}}\alpha\,,
\end{align}
\end{subequations}
in analogy to the transformation of a U(1) gauge field.
However,
integration of $\mathcal{A}_\psi$ around a closed curve 
$\mathcal{C}\subset\mathcal{P}_3$ 
generates a momentum-space geometric phase $\Phi_\mathcal{C}$ 
that is gauge invariant modulo $2\pi$, 
\begin{equation}
\label{eq:berry-phase}
\Phi_\mathcal{C} =
\oint_{\mathcal{C}} \mathrm{d}\mathbf{l}\,\mathcal{A}_{\psi}(\mathbf{p})\,,
\end{equation}
where $\mathrm{d}\mathbf{l}$ is the line element along $\mathcal{C}$. 
The phase $\Phi_\mathcal{C}$ is nonzero whenever the holonomy
of the curve $\mathcal{C}$ is nontrivial.
Another gauge-invariant quantity constructed from $\mathcal{A}_\psi$
is the Berry curvature 2-form 
\begin{equation}
\Omega_{\psi}(\mathbf{p})
:=\nabla_{\mathbf{p}}\times\mathcal{A}_{\psi}(\mathbf{p})\,.
\label{eq:berry-curvature-2}
\end{equation}
The analogy between $\mathcal{A}_\psi$
and the U(1) electromagnetic gauge potential 
implies that this curvature can be interpreted
as an effective magnetic field in momentum space.

For each fermion mode of interest here,
the dispersion $E=E(\mathbf{p})$ 
provides the eigenvalues of the hamiltonian for the quantum system. 
Extending momentum space $\mathcal{P}_3$ by the energy axis $p_0=E$ 
implies that the dispersion 
can be viewed as specifying a three-dimensional subspace 
of the four-dimensional energy-momentum space~$\mathcal{P}_4$. 
The set of all hamiltonian eigenstates 
$|\psi_E(\mathbf{p})\rangle$ then gives rise to a bundle 
$P({\rm U(1)},\mathcal{P}_3)$.
The various branches of the dispersion
can intersect or touch each other, 
indicating the presence of degeneracies. 
Degeneracies at an isolated point are particularly intriguing
because the Berry curvature diverges at the point, 
representing a situation analogous to that of a magnetic monopole
in electrodynamics. 
The corresponding connection possesses singularities 
along the Dirac string of the monopole. 
Consequently, 
no globally well-defined connection 
on $P({\rm U(1)},\mathcal{P}_3)$ exists.
The obstruction 
can be characterized by a topological charge called the first Chern number,
which can be computed by integrating the curvature 
over a closed surface $S$ containing the monopole,
\begin{equation}
\oint_S\mathrm{d}S\,\Omega_{\psi}(\mathbf{p})\cdot\hat{\mathbf{n}}=2\pi N_S\,.
\label{eq:berry-curvature-integrated}
\end{equation}
In this equation,
$\hat{\mathbf{n}}$ is a normal unit vector to $S$, 
$\mathrm{d}S$ is the surface element,
and $N_S\in\mathbb{Z}$ is the topological charge. 
It follows that the total Berry curvature flux 
through a closed surface is quantized in units of $2\pi$. 
In what follows,
we take advantage of these topological properties
to investigate the momentum-space geometric phases
associated with the fermion actions $S_b$ and $S_d$.

\section{CPT-odd case}
\label{sec:application-sme-b}

This section explores the momentum-space geometric phases 
associated with the action $S_b$.
We consider in turn scenarios with 
purely timelike, purely spacelike, and generic spacelike $b_\mu$ coefficients. 

In the massless limit,
the action $S_b$ implies the modified Dirac equation
\begin{equation}
(\gamma^{\mu}\lambda_{\mu}-b_{\mu}\gamma_5\gamma^{\mu})\psi=0\,,
\end{equation}
where $\lambda_{\mu}$ is the wave four-vector.
The Dirac equation can be written in Schr\"{o}dinger form
\begin{subequations}
\begin{align}
H_b\psi&=\lambda_0 \psi\,, \\[1ex]
\label{eq:dirac-hamiltonian-modified}
H_b&=H_0+b_{\mu}\gamma^0\gamma_5\gamma^{\mu}\,, \\[1ex]
\label{eq:dirac-hamiltonian-standard}
H_0&=-\gamma^0\gamma^i\lambda_i\,,
\end{align}
\end{subequations}
with a relativistic one-particle hamiltonian $H_b$ for the $b_{\mu}$ theory, which is a modification of the standard Dirac hamiltonian $H_0$. 

\subsection{Purely timelike coefficient}
\label{sec:purely-timelike-sector}

Consider a purely timelike $b_{\mu}$, which is isotropic, with $b_0>0$. According to Eq.~\eqref{eq:dirac-hamiltonian-modified}, the relativistic hamiltonian of a free fermion modified by $b_0$ is then of the form
\begin{equation}
\label{eq:dirac-hamiltonian-modified2}
H_{\ring{b}}=H_0-\ring{b}\gamma_5\,,
\end{equation}
where $\ring{b}=b_0$ denotes the isotropic piece of $b_{\mu}$~\cite{Kostelecky:2013rta}. In the following, all results are expressed in terms of the spatial-momentum 1-form $\mathcal{p}=(p_i)$ where $p:=|\mathcal{p}|=\sqrt{\sum_i p_i^2}$.

Inspection reveals that $[H_0,H_{\ring{b}}]=0$, 
which implies that a common set of eigenvectors can be chosen for 
the standard Dirac hamiltonian and the isotropic modified hamiltonian.
Like $H_0$, $H_{\ring{b}}$ exhibits negative eigenvalues. 
In the presence of $b_0$,
however, 
the energy eigenvalues lie on Weyl cones shifted by $b_0$ along the positive and negative energy axes, respectively. The parts of these double cones below the Weyl points have either completely or primarily negative eigenvalues. The usual reinterpretation procedure reflects each part to the corresponding cone above the Weyl nodes. This yields 
\begin{subequations}
\label{eq:energy-eigenvalues}
\begin{align}
E_u^{(1,2)}&=p\pm b_0\,, \\[1ex]
E_v^{(1,2)}&=p\mp b_0\,.
\end{align}
\end{subequations}
Here and in what follows, 
the labels (1) and (2) are correlated with the signs on the right-hand sides. 
Note that $E_u^{(2)}$ and $E_v^{(1)}$ dip below the $E=0$ plane, 
whereupon negative energies are an inevitable consequence.
This feature is one aspect of the concordance problem,
for which a physical resolution is discussed
in Ref.~\cite{AlanKostelecky:2024gek}.

\begin{figure}
\centering
\subfloat[]{\label{fig:dispersions-timelike-b}\includegraphics[scale=0.2]{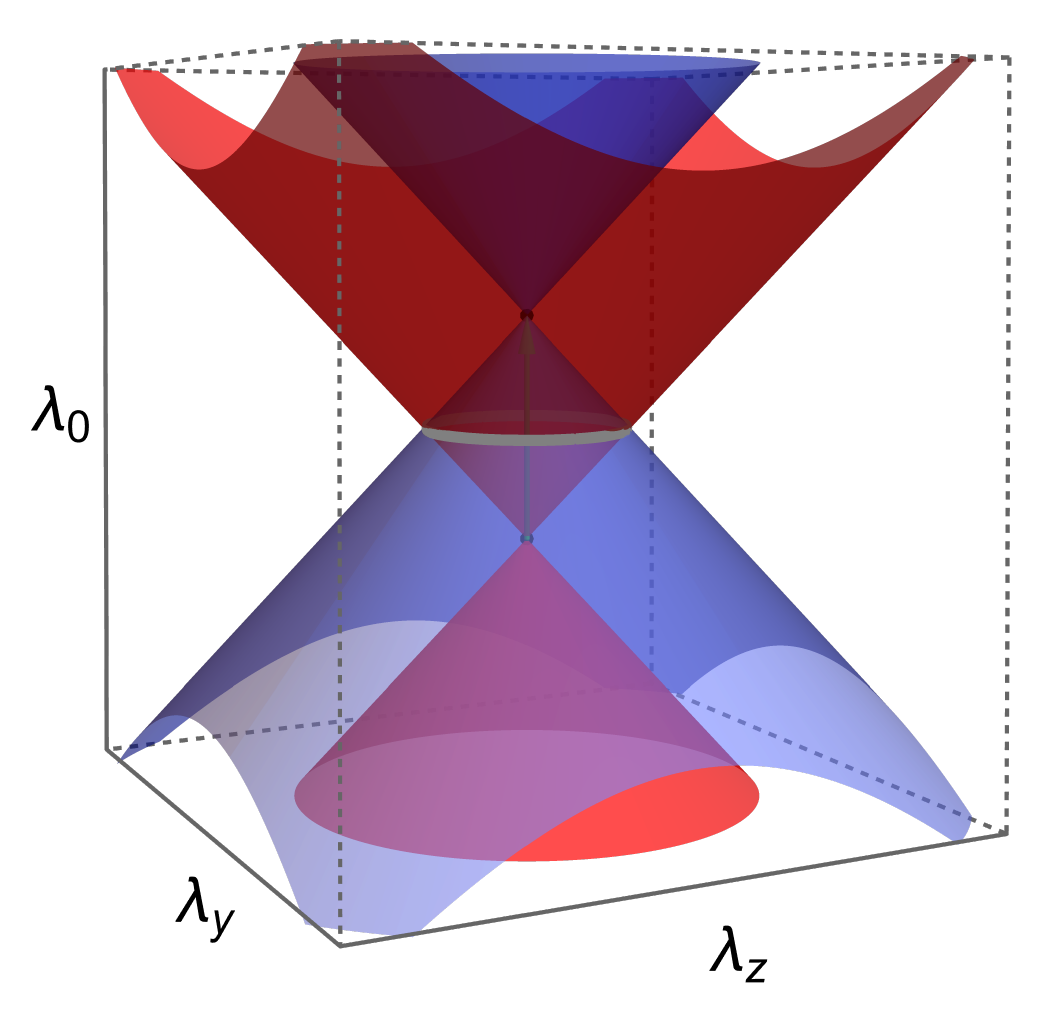}}
\subfloat[]{\label{fig:dispersions-timelike-b-reinterpreted}\includegraphics[scale=0.2]{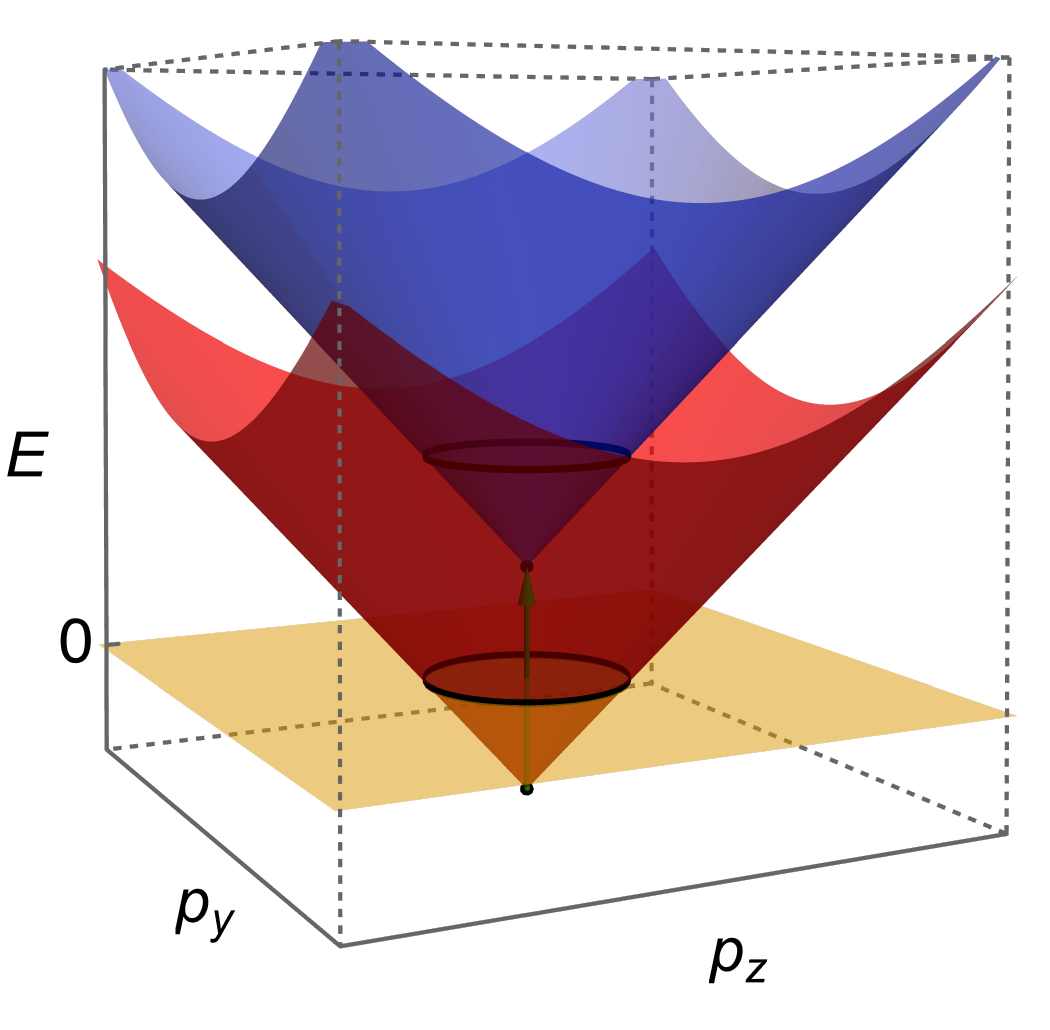}}
\caption{Energy eigenvalues for the timelike case
\protect\subref{fig:dispersions-timelike-b}~before and \protect\subref{fig:dispersions-timelike-b-reinterpreted} after reinterpretation.}
\label{fig:dispersions-timelike-b-total}
\end{figure}

The energy eigenvalues 
of Eq.~\eqref{eq:dirac-hamiltonian-modified} 
are illustrated in Fig.~\ref{fig:dispersions-timelike-b-total}. 
In these plots,
$\lambda_{\mu}$ is the wave 4-vector of a generic plane wave, 
whereas $p_{\mu}=(E,\mathcal{p})$ 
denotes the physical 4-momentum after reinterpretation. 
The yellow plane illustrates the Fermi level,
and the black circles represent loops 
along which energy eigenstates can be parallel transported 
around the apices of the Weyl cones.

Applying the usual reinterpretation, the four spinor-valued energy eigenstates as functions of cartesian momentum-space coordinates can be cast into the form
\begin{subequations}
\label{eq:energy-eigenstates-timelike}
\begin{align}
|u^{(1,2)}(\mathcal{p})\rangle&=|v^{(1,2)}(\mathcal{p})\rangle=\mathcal{N}^{(1,2)}\begin{pmatrix}
\pm (p_x-\mathrm{i}p_y) \\
p\mp p_z \\
-(p_x-\mathrm{i}p_y) \\
\mp p+p_z \\
\end{pmatrix}\,,
\end{align}
with the normalization factors
\begin{equation}
\mathcal{N}^{(1,2)}=\frac{1}{2\sqrt{p(p\mp p_z)}}\,,
\end{equation}
\end{subequations}
such that
\begin{equation}
\langle u^{(1,2)}|u^{(1,2)}\rangle=1=\langle v^{(1,2)}|v^{(1,2)}\rangle\,.
\end{equation}
\begin{figure}
\centering
\subfloat[]{\label{fig:vector-potential-dirac-string}\includegraphics[scale=0.2]{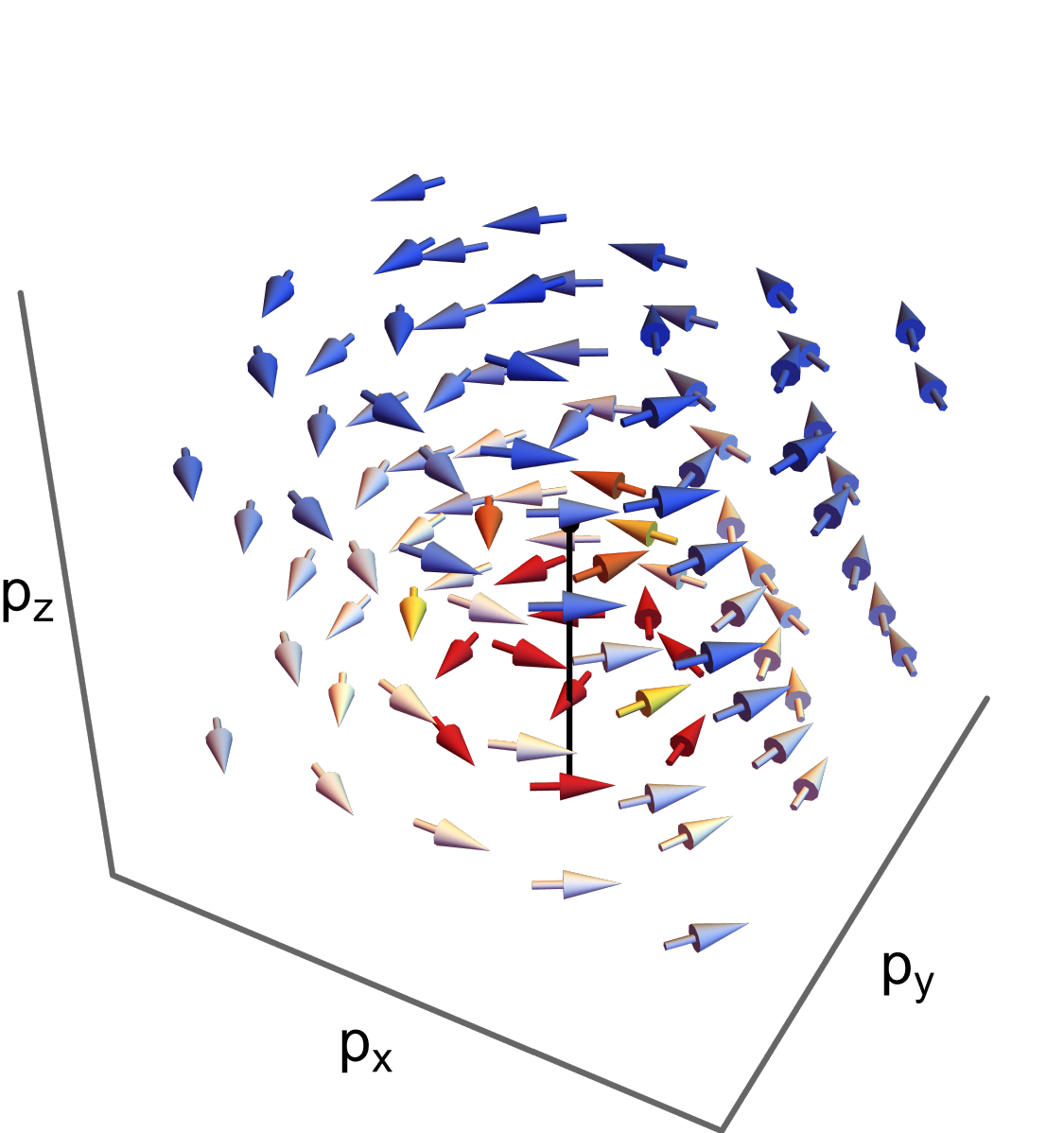}}\hspace{0.2cm}
\subfloat[]{\label{fig:dirac-string-contour-plot}\includegraphics[scale=0.13]{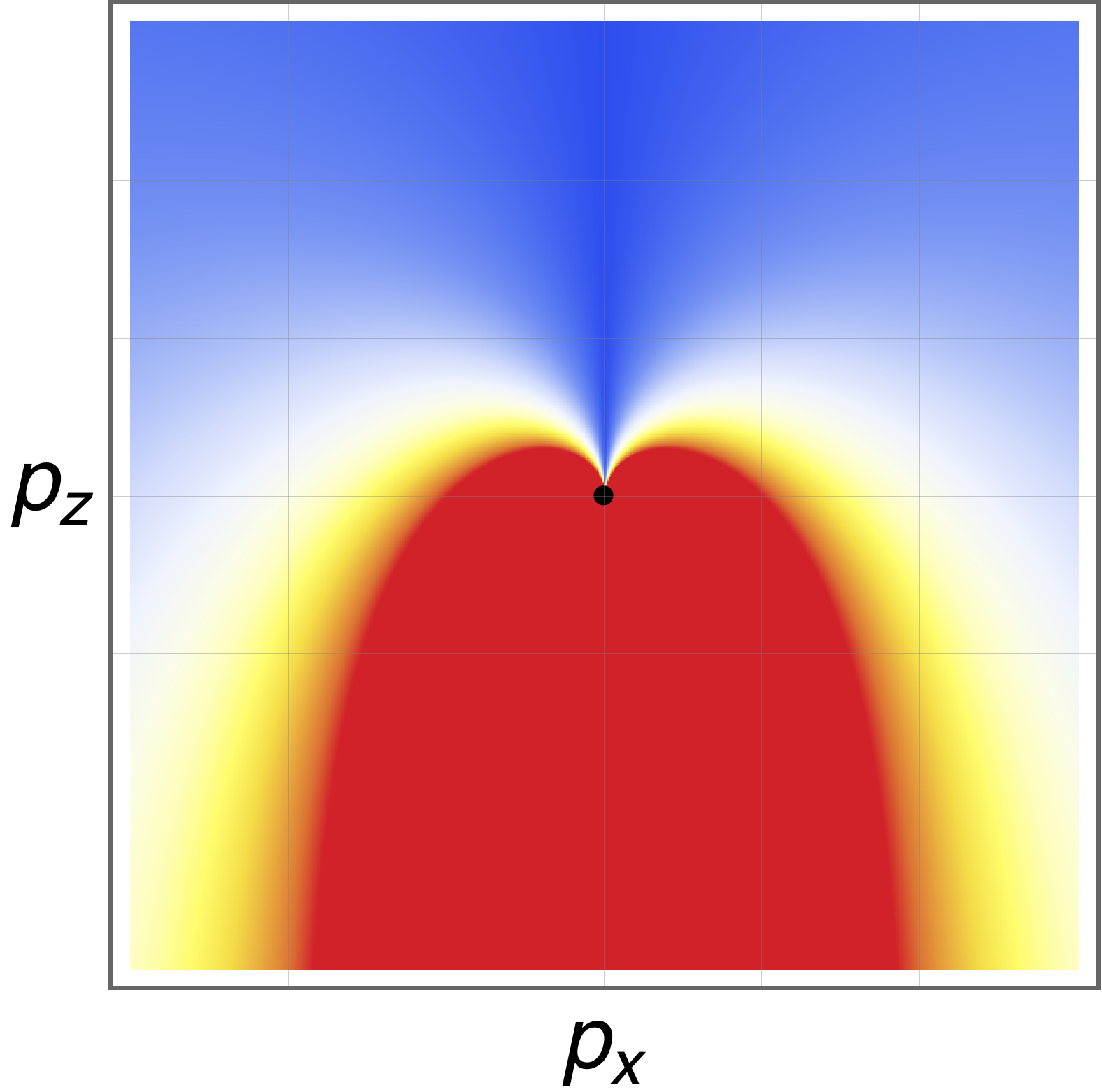}}
\caption{
\protect\subref{fig:vector-potential-dirac-string} 
Berry connection vector field~\eqref{eq:berry-connection-u-2},
\quad\quad
and \protect\subref{fig:dirac-string-contour-plot} 
contour plot of its norm.}
\label{fig:dirac-string}
\end{figure}
An alternative way of expressing these eigenstates is via spherical momentum-space coordinates $(p,\theta,\phi)$, where $\theta$ is the polar angle and $\phi$ is the azimuthal angle,
\begin{subequations}
\label{eq:energy-eigenstates-spherical-coordinates}
\begin{align}
|u^{(1)}(\mathcal{p})\rangle&=\frac{\exp(-\mathrm{i}\phi)}{\sqrt{2}}\begin{pmatrix}
\xi^{(1)} \\
-\xi^{(1)} \\
\end{pmatrix}=|v^{(1)}(\mathcal{p})\rangle\,, \displaybreak[0]\\[2ex]
|u^{(2)}(\mathcal{p})\rangle&=\frac{1}{\sqrt{2}}\begin{pmatrix}
\xi^{(2)} \\
\xi^{(2)} \\
\end{pmatrix}=|v^{(2)}(\mathcal{p})\rangle\,, 
\end{align}
\end{subequations}
with the auxiliary two-component spinors
\begin{subequations}
\label{eq:auxiliary-spinors}
\begin{align}
\xi^{(1)}&=\xi^{(1)}(\theta,\phi)=\begin{pmatrix}
\cos(\theta/2) \\
\sin(\theta/2)\exp(\mathrm{i}\phi) \\
\end{pmatrix}\,, \\[2ex]
\xi^{(2)}&=\xi^{(2)}(\theta,\phi)=\begin{pmatrix}
-\sin(\theta/2)\exp(-\mathrm{i}\phi) \\
\cos(\theta/2) \\
\end{pmatrix}\,.
\end{align}
\end{subequations}
From the latter form we deduce that Eq.~\eqref{eq:energy-eigenstates-timelike} represents helicity eigenstates. They correspond to the spinor solutions obtained in Ref.~\cite{Colladay:1996iz},
modulo global phases. 

In evaluating the momentum-space geometric phase 
using Eq.~\eqref{eq:berry-phase}, 
the Dirac string is best revealed in spherical coordinates via Eqs.~\eqref{eq:energy-eigenstates-spherical-coordinates}, \eqref{eq:auxiliary-spinors}. The eigenstates $|u^{(1)}\rangle$ and $|v^{(1)}\rangle$ have a $\phi$-dependent limit when the negative $p_z$ axis is approached, i.e., for $\theta\rightarrow\pi$. Thus, the limit of these states along the negative $p_z$ axis is undefined. A similar behavior is observed for $|u^{(2)}\rangle$ and $|v^{(2)}\rangle$, but when we move toward the positive $p_z$ axis, i.e., for $\theta\rightarrow 0$. This behavior is gauge dependent and changes under a gauge transformation.
In cartesian and spherical coordinates, respectively,
the Berry connection becomes  
\begin{subequations}
\label{eq:berry-connection-u}
\begin{align}
\label{eq:berry-connection-u-1}
\mathcal{A}^{(1)}_u&=\frac{1}{2p(p-p_z)}\begin{pmatrix}
-p_y \\
p_x \\
0 \\
\end{pmatrix}=\frac{1}{2p}\cot\left(\frac{\theta}{2}\right)\hat{\phi}=\mathcal{A}^{(1)}_v\,, \displaybreak[0]\\
\label{eq:berry-connection-u-2}
\mathcal{A}^{(2)}_u&=\frac{1}{2p(p+p_z)}\begin{pmatrix}
-p_y \\
p_x \\
0 \\
\end{pmatrix}=\frac{1}{2p}\tan\left(\frac{\theta}{2}\right)\hat{\phi}=\mathcal{A}^{(2)}_v\,,
\end{align}
\end{subequations}
with the unit vector  $\hat{\phi}$ tangential to circles of latitude of the sphere. The Dirac strings for $\mathcal{A}^{(1)}_{u,v}$ and $\mathcal{A}^{(2)}_{u,v}$ are evident when $\theta\mapsto 0$ and $\theta\mapsto \pi$, respectively.
Figure~\ref{fig:dirac-string} illustrates the second case,
with the temperature coloring indicating the Dirac string
along the negative $p_z$ axis.
Note that the $\mathcal{A}^{(1,2)}_u$ in Eq.~\eqref{eq:berry-connection-u} correspond to the gauge potentials chosen in the Wu-Yang construction for a magnetic monopole~\cite{Wu:1975es}, up to global phases.

\begin{figure}
\centering
\subfloat[]{\label{fig:magnetic-monopole-arrows}\includegraphics[scale=0.2]{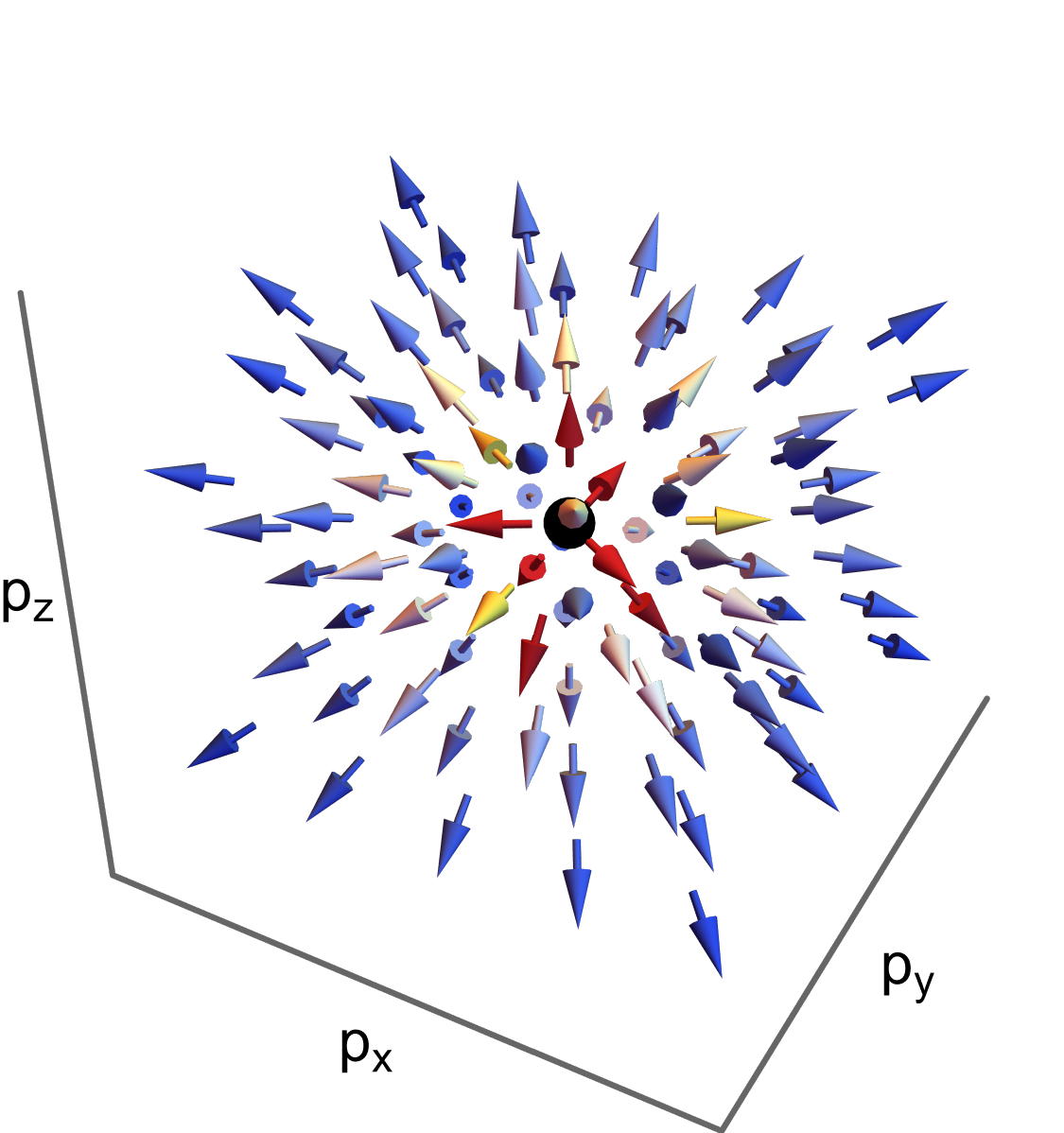}}\hspace{0.2cm}
\subfloat[]{\label{fig:magnetic-monopole-contour-plot}\includegraphics[scale=0.13]{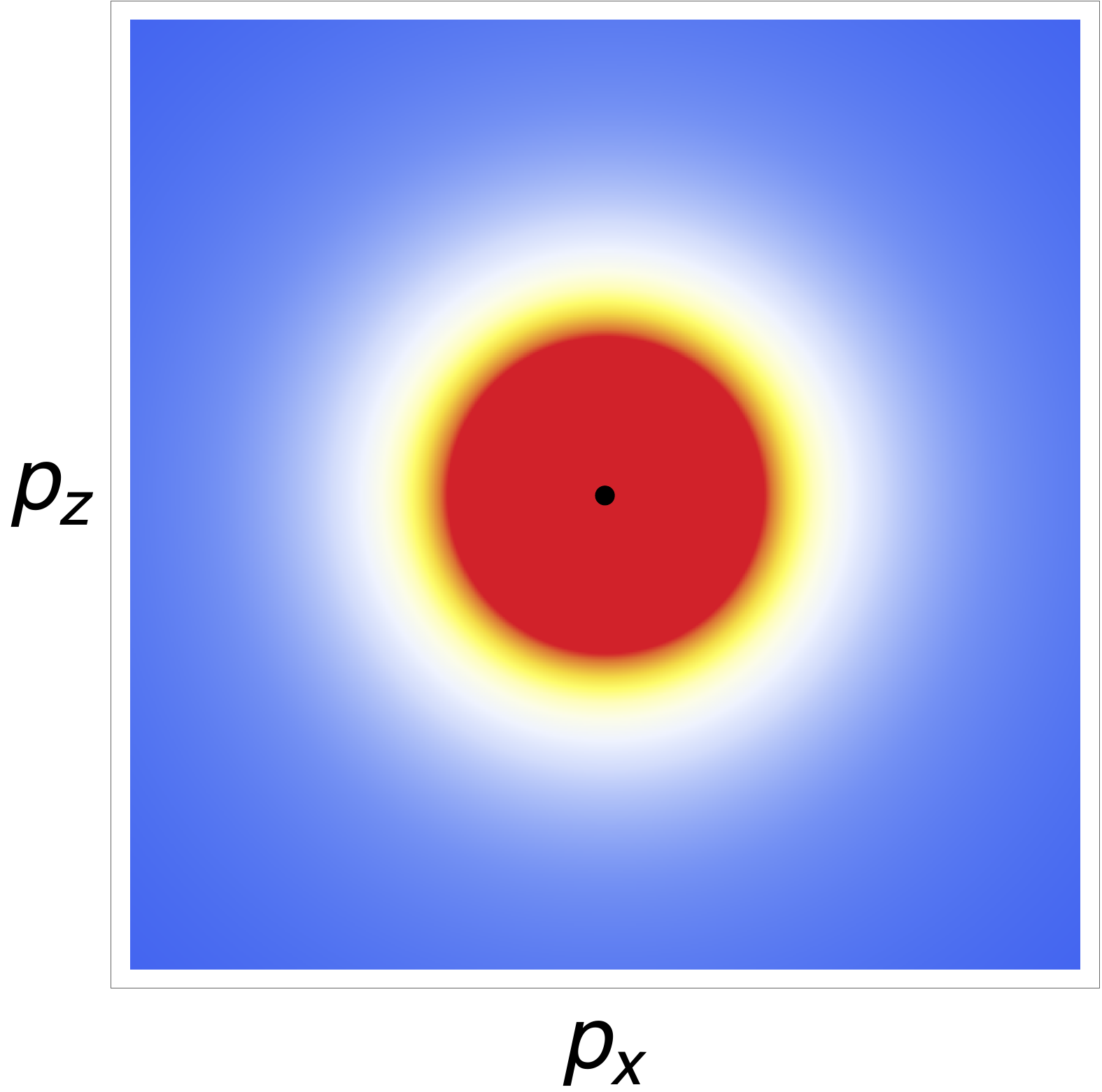}}
\caption{
\protect\subref{fig:magnetic-monopole-arrows} 
Vector field from a source of Berry curvature 
in Eq.~\eqref{eq:berry-curvature-isotropic-b},
and \protect\subref{fig:magnetic-monopole-contour-plot} 
contour plot of its norm.}
\label{fig:magnetic-monopole}
\end{figure}

The curvature of Eq.~\eqref{eq:berry-curvature-2} is explicitly evaluated for the eigenstates of Eq.~\eqref{eq:energy-eigenstates-timelike} to be
\begin{equation}
\label{eq:berry-curvature-isotropic-b}
\Omega^{(1,2)}_{u,i}=\mp\frac{p_i}{2p^3}=\Omega^{(1,2)}_{v,i}\,.
\end{equation}
Hence, nonzero contributions with opposite signs arise at the origin of momentum space.
Each of these has the form of a magnetic monopole,
illustrated in Fig.~\ref{fig:magnetic-monopole}. 
The temperature coloring reveals the increasing curvature 
approaching the origin,
and the monopole is indicated by a black dot.
Each monopole has strength $g=1/2$, 
half that of the Wu-Yang monopole \cite{Wu:1975es}.

To determine the first Chern numbers associated with the monopoles, we integrate the curvature over a suitable closed surface containing the singularities according to Eq.~\eqref{eq:berry-curvature-integrated}. It is convenient to choose a 2-sphere $S^2_{\mathcal{p}_0}$ of radius $p$ and normal $\mathcal{n}$ around a point $\mathcal{p}_0\in\mathcal{P}_3$. Here, we pick the origin of momentum space such that $\mathcal{p}_0=0$. Then,
\begin{equation}
\label{eq:berry-connection-integrated}
\int_{S^2_{\mathcal{p}_0}}\mathrm{d}\Omega\,p^2\Omega^{(1,2)}_u\cdot\mathcal{n}=\mp 2\pi=\int_{S^2_{\mathcal{p}_0}}\mathrm{d}\Omega\,p^2\Omega^{(1,2)}_v\cdot\mathcal{n}\,,
\end{equation}
where the unit normal $\mathcal{n}$ to the surface is pointing outwards. 
This reveals that the first Chern numbers are
\begin{equation}
N_u^{(1,2)}= N_v^{(1,2)}= \mp 1 
\quad {\rm (purely~timelike~case)} \,.
\end{equation}

The results~\eqref{eq:berry-curvature-isotropic-b} are independent of $b_0$, since the eigenstates of Eq.~\eqref{eq:energy-eigenstates-timelike} correspond to the Lorentz-invariant ones due to $[H_0,H_{\ring{b}}]=0$. In contrast, the dispersions~\eqref{eq:energy-eigenvalues} do depend on $b_0$. The curvature characterizes the geometry of $P({\rm U(1)},\mathcal{P}_3)$ and hence is sensitive to the Weyl nodes at $\mathcal{p}=0$. However, it is insensitive to shifts of the nodes by $\pm b_0$ along the energy axis. Furthermore, as the curvature only detects topologically protected isolated points, the nodal surface $p=b_0$ plays no direct role.

Gauge transformations can be performed on the energy eigenstates such that $|u^{(1,2)}\rangle\mapsto |u^{(1,2)}\rangle'=\exp(\mathrm{i}\phi)|u^{(1,2)}\rangle$ and $|v^{(1,2)}\rangle\mapsto |v^{(1,2)}\rangle'=\exp(\mathrm{i}\phi)|v^{(1,2)}\rangle$. The resulting Berry connections are then given by
\begin{subequations}
\begin{align}
\mathcal{A}^{(1)}_{u'}&=-\frac{1}{2p}\tan\left(\frac{\theta}{2}\right)\hat{\phi}=\mathcal{A}^{(1)}_{v'}\,, \\[1ex]
\mathcal{A}^{(2)}_{u'}&=-\frac{1}{2p}\cot\left(\frac{\theta}{2}\right)\hat{\phi}=\mathcal{A}^{(2)}_{v'}\,.
\end{align}
\end{subequations}
The gauge transformation preserves the Dirac strings while changing their location in momentum space: $\mathcal{A}^{(1)}_{u',v'}$ and $\mathcal{A}^{(2)}_{u',v'}$ have singularities for $\theta\mapsto \pi$ and $\theta\mapsto 0$, respectively. 

\begin{figure}
\centering
\includegraphics[scale=0.4]{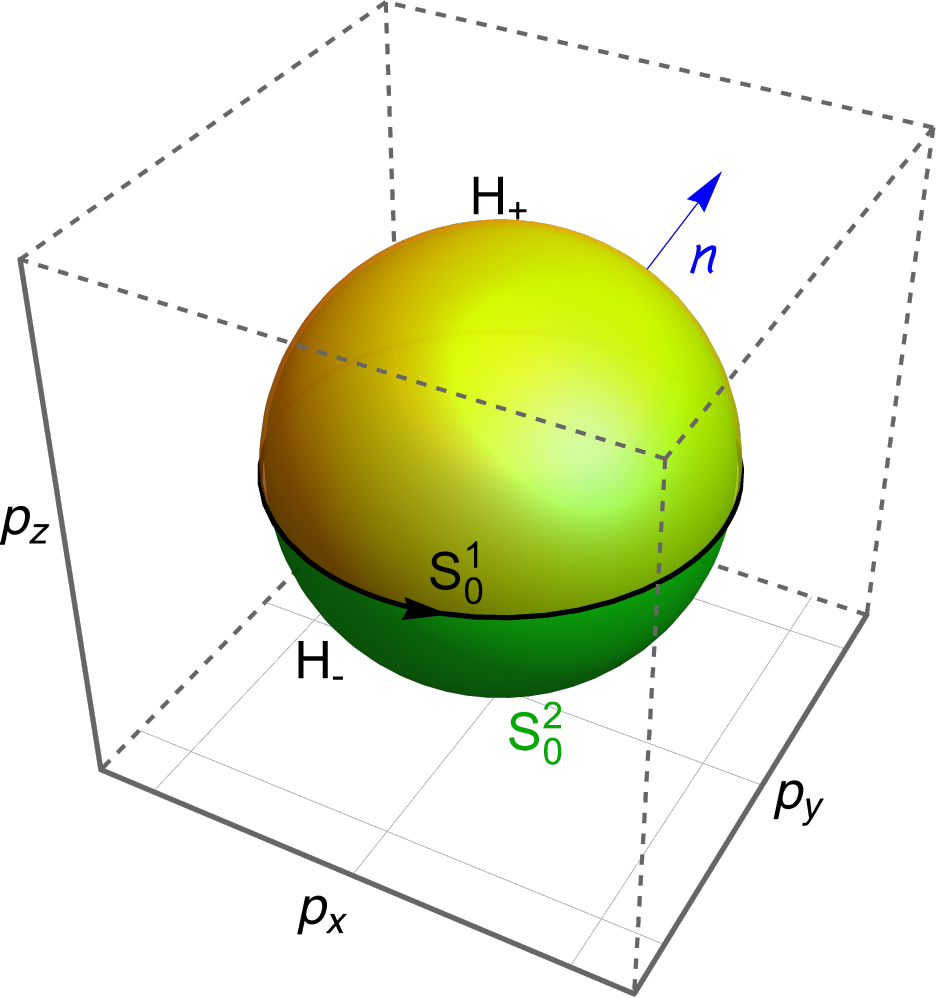}
\caption{2-sphere $S_{0}^2$ around magnetic monopole with unit normal $\mathcal{n}$ and upper hemisphere $H_+$ bounded by the circle $S_{0}^1$.}
\label{fig:integrations}
\end{figure}

Next, we determine the momentum-space geometric phase accumulated along a circle around the apex of a Weyl cone. Consider first a circle $S^1_{0}$ in the $p_x$-$p_y$ plane, which is centered at $p_i=0$ and is to be integrated in counterclockwise direction. 
Note that the Dirac strings of the eigenstates $|u^{(1)}\rangle$ and $|u^{(2)}\rangle$ are located at $p_z\in (0,\infty)$ and $p_z\in (-\infty,0)$, respectively, such that $p_x=p_y=0$,
while the Dirac strings of $|v^{(1)}\rangle$ and $|v^{(2)}\rangle$ are at $p_z\in (0,\infty)$ and $p_z\in (-\infty,0)$, respectively, with $p_x=p_y=0$. So the Dirac strings lie outside $S^1_{0}$,
and the integrations can be carried out without issues.
Using the Stokes theorem,
the line integral of Eq.~\eqref{eq:berry-phase} can be reformulated as a surface integral over the curvature 2-form $\Omega_{\psi}$, 
\begin{align}
\label{eq:berry-phase-surface-integral}
\Phi&=\int_{H^+} \mathrm{d}\Omega\,p^2[\nabla_{\mathbf{p}}\times\mathcal{A}_{\psi}(\mathbf{p})]\cdot \mathcal{n} \notag \\
&=\int_{H^+} \mathrm{d}\Omega\,p^2\Omega_{\psi}(-\mathcal{p})\cdot \mathcal{n}\,.
\end{align}
Note that $H^+$ corresponds to the upper hemisphere of $S^2_{0}$ such that $\partial H^+=S^1_{0}$,
as illustrated in Fig.~\ref{fig:integrations}. Hence, the surface that is integrated over is a subset of the surface considered in Eq.~\eqref{eq:berry-connection-integrated}. The value of $\Phi$ in Eq.~\eqref{eq:berry-phase-surface-integral} corresponds to the Berry flux penetrating the surface $H^+$.

Recall that the Stokes theorem is only applicable when neither $H^+$ nor $\partial H^+$ contain singularities. However, the former is clearly not the case for $|u^{(1)}\rangle$ and $|v^{(1)}\rangle$ due to their Dirac strings along the positive $p_z$ axis. Thus, in this case we have to integrate along $S^1_0$ in clockwise direction, with the corresponding surface as the lower hemisphere $H^-$ of $S^2_{0}$. Based on Eq.~\eqref{eq:berry-curvature-isotropic-b}, the opposite sign of the curvature $\Omega_{u,v}^{(1)}$ as compared to $\Omega_{u,v}^{(2)}$ comes into play, whereas the areas of $H^{\pm}$ are equal. Therefore, performing either the line integrals along $S^1_{0}$ in the counterclockwise (clockwise) direction or the surface integral of Eq.~\eqref{eq:berry-phase-surface-integral} over $H^+$ ($H^-$) leads 
to the momentum-space geometric phases
\begin{equation}
\Phi^{(1)}_u=-\pi=\Phi^{(1)}_v\,,\quad \Phi^{(2)}_u=\pi=\Phi^{(2)}_v\,.
\label{eq:berry-phases-timelike-case}
\end{equation}

We can similarly determine the momentum-space geometric phases around a circle $S^1_{0}$ situated in the $p_x$-$p_z$ plane, i.e., for $\phi=\pi/2$,
as illustrated in~Fig.~\ref{fig:dispersions-timelike-b-reinterpreted}. This situation requires handling the Dirac strings along the $p_z$ axis,
using a gauge transformation to remove them. One possibility is to apply the following {\rm U(1)} transformations on the $u$-type energy eigenstates of Eq.~\eqref{eq:energy-eigenstates-timelike} with label $s=1,2$:
\begin{subequations}
\label{eq:gauge-transformation-dirac-string}
\begin{align}
|u^{(s)}(\mathcal{p})\rangle&\mapsto |u^{(s)}(\mathcal{p})\rangle'=\exp(\mathrm{i}\chi^{(s)})|u^{(s)}(\mathcal{p})\rangle\,, \displaybreak[0]\\[1ex]
\chi^{(1)}&=-\arctan\left[\cot(\phi)-\tan\left(\frac{\theta}{2}\right)\csc(\phi)\right]\,, \displaybreak[0]\\[1ex]
\chi^{(2)}&=\phi+\arctan\left[\cot(\phi)+\tan\left(\frac{\theta}{2}\right)\csc(\phi)\right]\,.
\end{align}
\end{subequations}
The $v$-type eigenstates of Eq.~\eqref{eq:energy-eigenstates-timelike} should be gauge-transformed analogously. By doing so, the Dirac strings are moved to straight lines along the $p_x$ axis,
with those for $|u^{(1)}\rangle'$, $|v^{(1)}\rangle'$ aligned with $p_x\in (0,\infty)$, $p_y=p_z=0$ and those for $|u^{(2)}\rangle'$, $|v^{(2)}\rangle'$ situated at $p_x\in (-\infty,0)$, $p_y=p_z=0$. The transformed situation is then analogous to the setting in the $p_x$-$p_y$ plane considered previously, thereby reproducing the result~\eqref{eq:berry-phases-timelike-case}.

\subsection{Purely spacelike coefficient}
\label{sec:purely-spacelike-b}

\begin{figure}
\centering
\subfloat[]{\label{fig:dispersions-purely-spacelike}\includegraphics[scale=0.2]{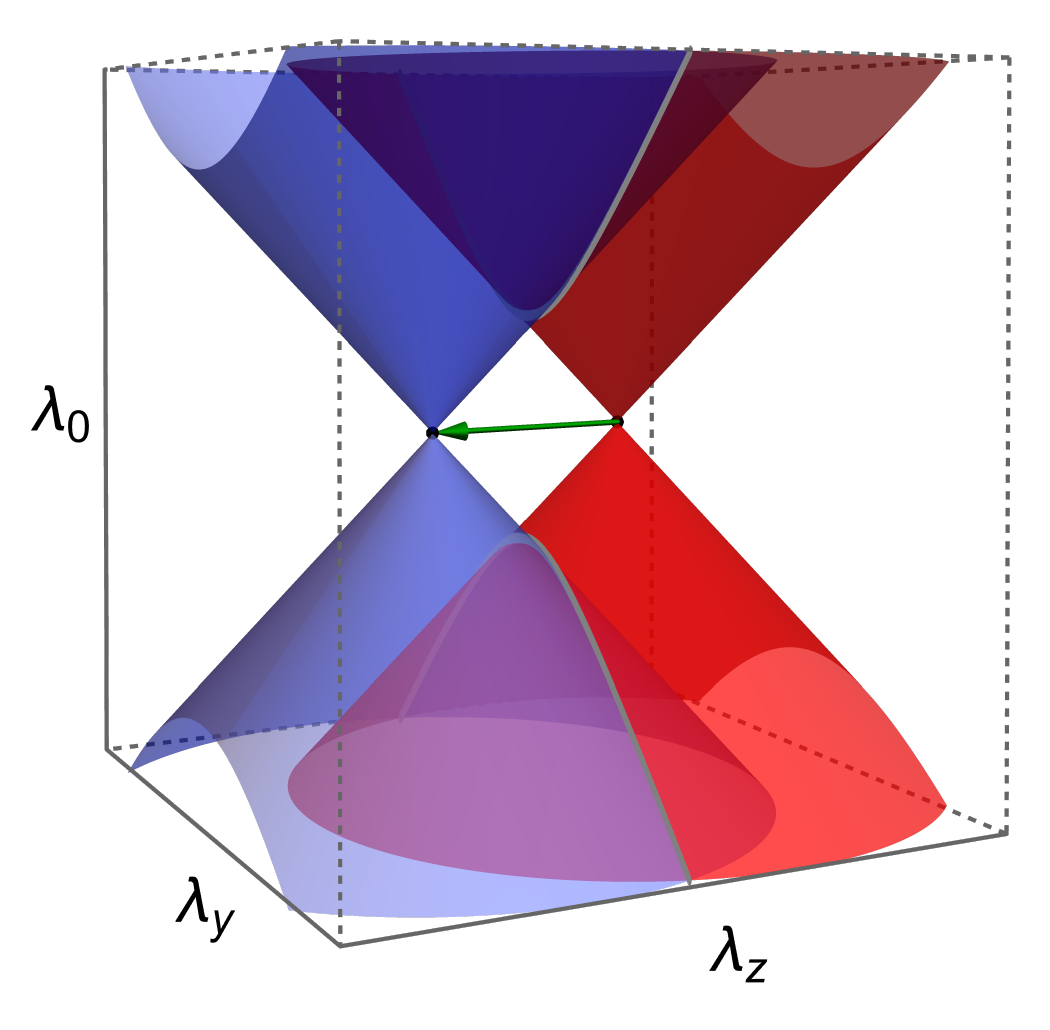}}
\subfloat[]{\label{fig:dispersions-purely-spacelike-reinterpreted}\includegraphics[scale=0.2]{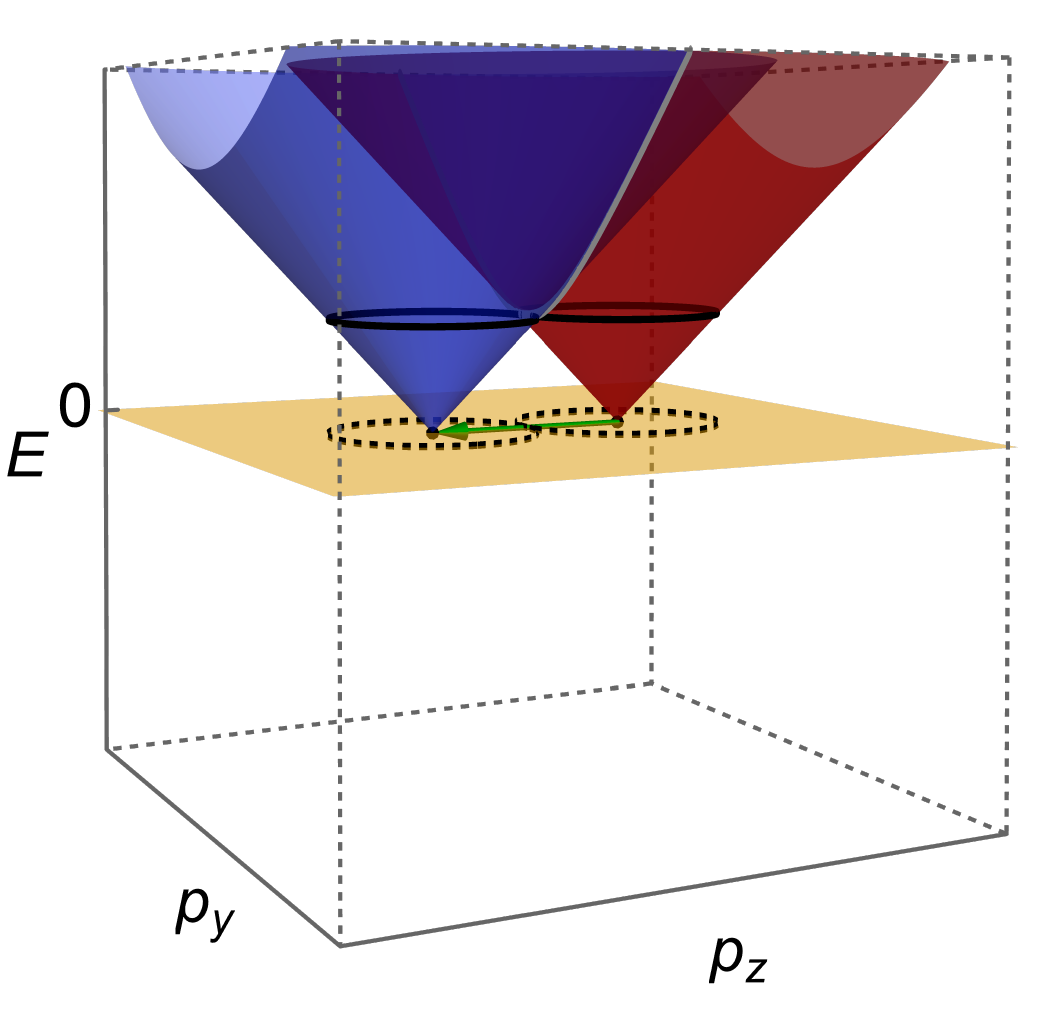}}
\caption{Energy eigenvalues for the spacelike case
\protect\subref{fig:dispersions-purely-spacelike}~before and \protect\subref{fig:dispersions-purely-spacelike-reinterpreted} after reinterpretation.}
\label{fig:dispersions-purely-spacelike-total}
\end{figure}

We next consider the case of a purely spacelike coefficient $b_\mu$, 
which incorporates spatial anisotropy. 
Without loss of generality,
we can take $\mathcal{b}=(0,0,b_z)$ and $b_z>0$.
The hamiltonian then reads
\begin{equation}
\label{eq:hamiltonian-purely-spacelike}
H_{\mathcal{b}}=H_0+b_z\gamma^0\gamma_5\gamma^z\,,
\end{equation}
with the standard Dirac hamiltonian $H_0$ of Eq.~\eqref{eq:dirac-hamiltonian-standard}. Note that $[H_0,H_{\mathcal{b}}]\neq 0$, in contrast to the isotropic sector considered previously. The energy eigenvalues lie on Weyl cones shifted along the positive or negative $p_z$ axis by $b_z$,
as displayed in Fig.~\ref{fig:dispersions-purely-spacelike}. Applying the standard reinterpretation procedure generates two Weyl cones with apices on the $E=0$ plane and with all other states having positive energy, as illustrated in Fig.~\ref{fig:dispersions-purely-spacelike-reinterpreted}. This gives 
\begin{subequations}
\label{eq:dispersions-spacelike-b}
\begin{align}
E_u^{(1,2)}&=|\mathcal{p}\mp\mathcal{b}|\,, \\[1ex]
E_v^{(1,2)}&=|\mathcal{p}\pm\mathcal{b}|\,.
\end{align}
\end{subequations}

After reinterpretation, the energy eigenstates as functions of cartesian momentum-space coordinates are readily obtained as
\begin{subequations}
\label{eq:energy-eigenstates-purely-spacelike-case}
\begin{align}
|u^{(1,2)}(p_i)\rangle&=\mathcal{N}_u^{(1,2)}\begin{pmatrix}
\pm(p_x-\mathrm{i}p_y) \\
E_u^{(1,2)}\mp p_z+b_z \\
-(p_x-\mathrm{i}p_y) \\
\mp E_u^{(1,2)}+p_z\mp b_z \\
\end{pmatrix}\,, \displaybreak[0]\\[2ex]
|v^{(1,2)}(p_i)\rangle&=\mathcal{N}_v^{(1,2)}\begin{pmatrix}
\pm(p_x-\mathrm{i}p_y) \\
E_v^{(1,2)}\mp p_z-b_z \\
-(p_x-\mathrm{i}p_y) \\
\mp E_v^{(1,2)}+p_z\pm b_z \\
\end{pmatrix}\,,
\end{align}
\end{subequations}
with the normalization factors
\begin{subequations}
\begin{align}
\mathcal{N}_u^{(1,2)}&=\frac{1}{2\sqrt{E_u^{(1,2)}(E_u^{(1,2)}\mp p_z+b_z)}}\,, \\[1ex]
\mathcal{N}_v^{(1,2)}&=\frac{1}{2\sqrt{E_v^{(1,2)}(E_v^{(1,2)}\mp p_z-b_z)}}\,.
\end{align}
\end{subequations}

\begin{figure*}
\centering
\subfloat[]{\label{fig:magnetic-monopoles-purely-spacelike}\includegraphics[scale=0.2]{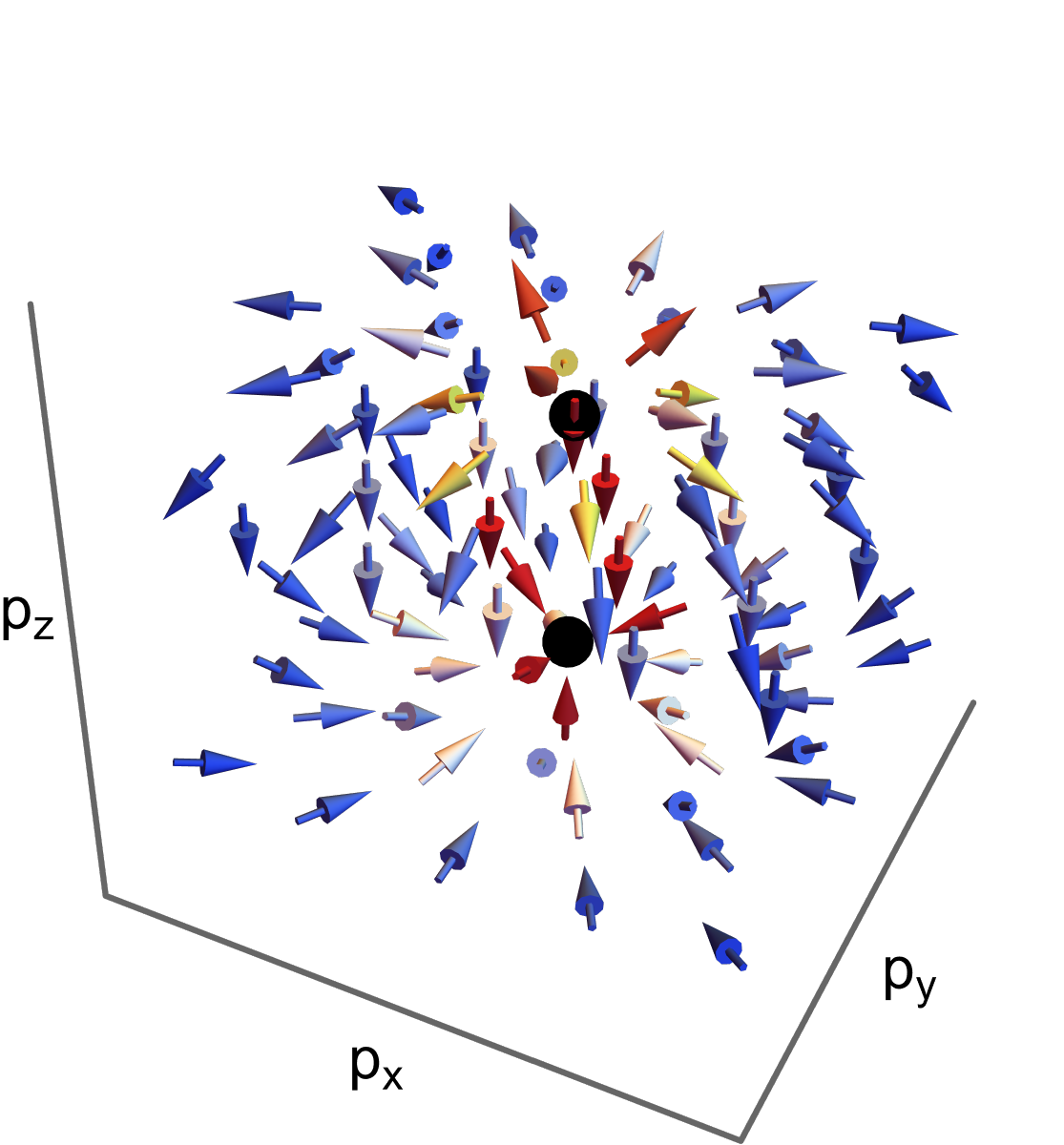}}\hspace{1cm}
\subfloat[]{\label{fig:berry-curvature-b-contour-plot-x-y}\includegraphics[scale=0.13]{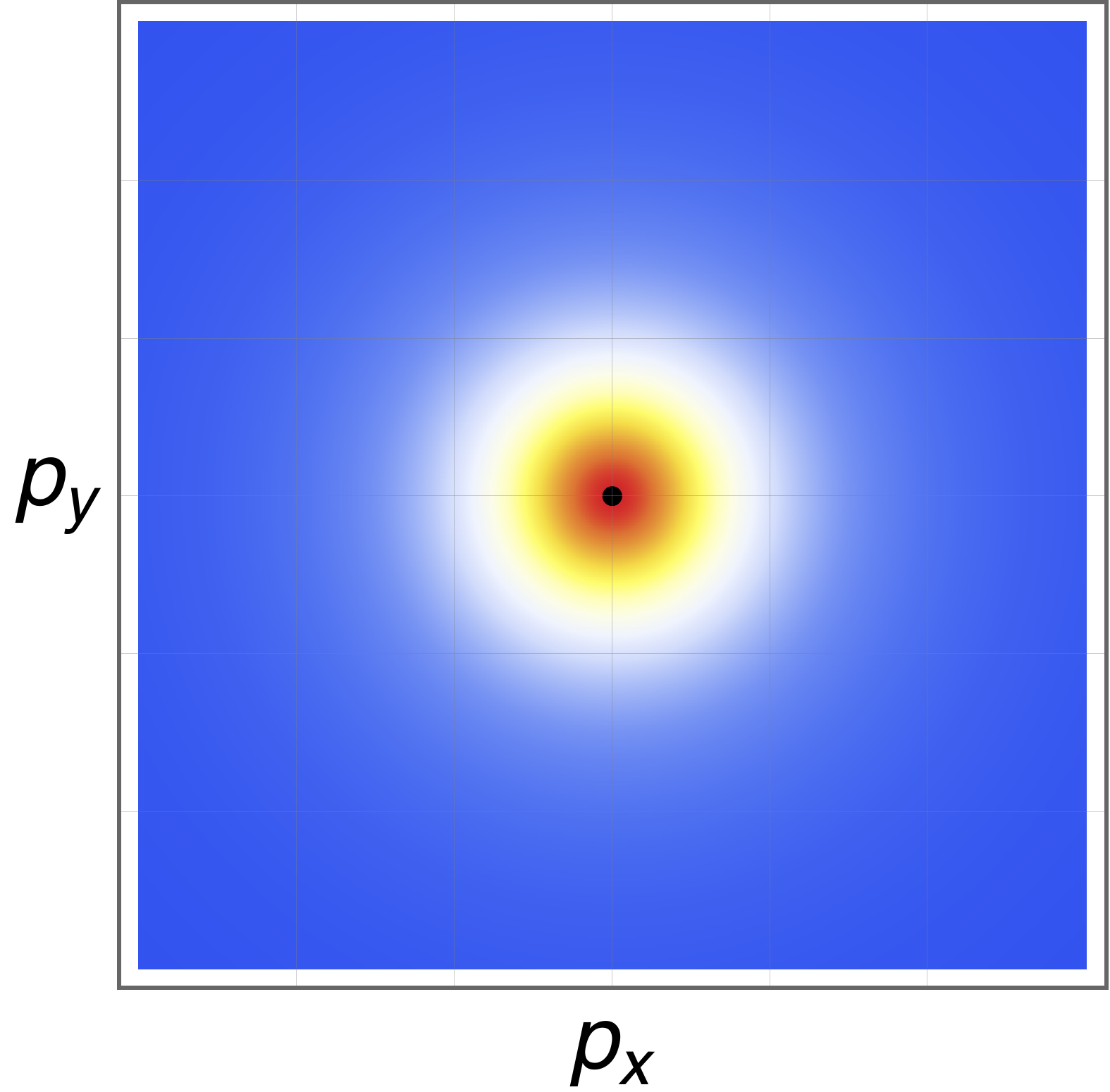}}\hspace{1cm}
\subfloat[]{\label{fig:berry-curvature-b-contour-plot-y-z}\includegraphics[scale=0.13]{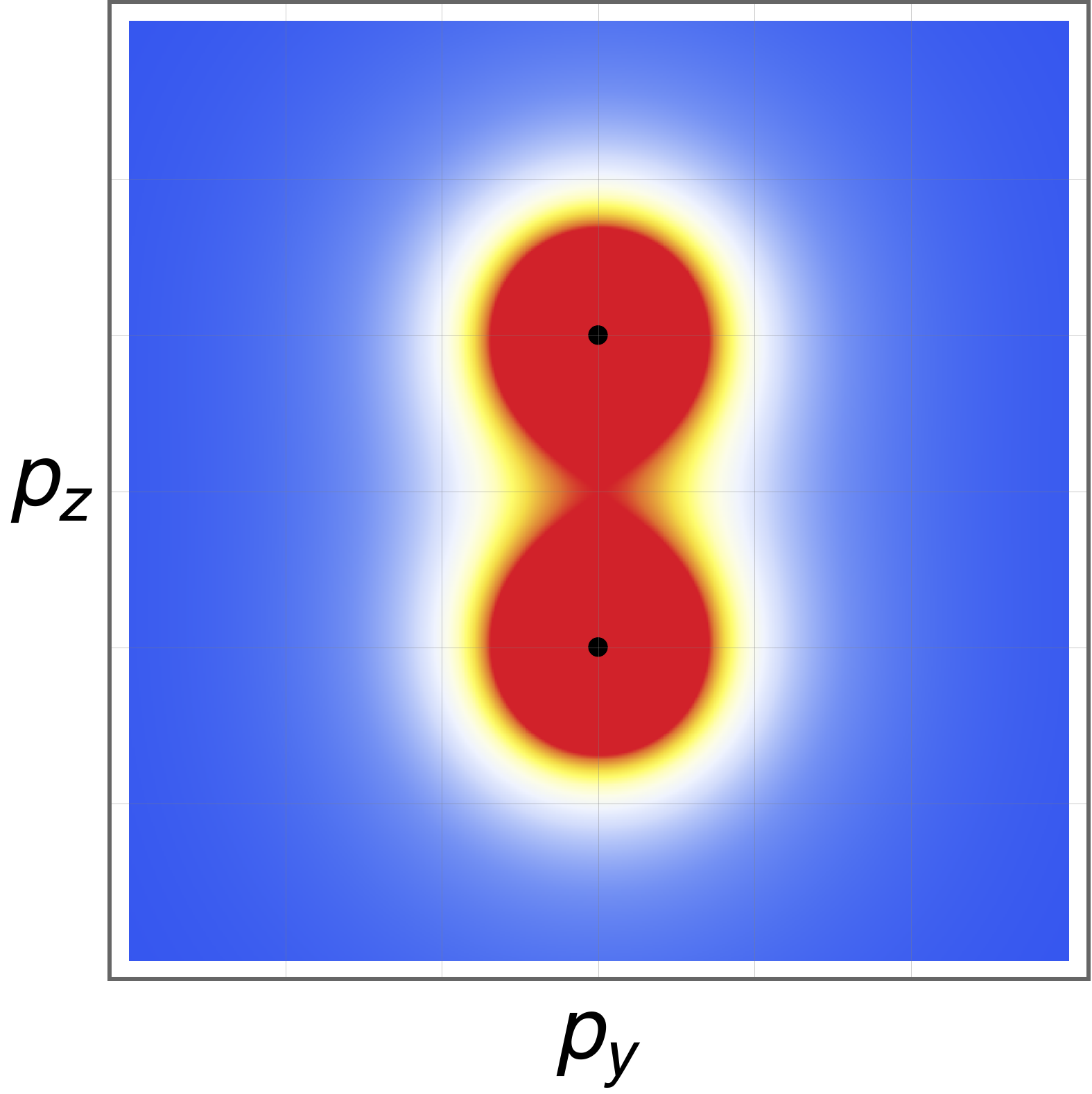}}
\caption{
\protect\subref{fig:magnetic-monopoles-purely-spacelike}
Vector-field plot of the curvature~\eqref{eq:berry-curvature-purely-spacelike-b},
\protect\subref{fig:berry-curvature-b-contour-plot-x-y}
contour plot of its norm in the $p_x$-$p_y$ plane,
and 
\protect\subref{fig:berry-curvature-b-contour-plot-y-z}
contour plot of its norm in the $p_y$-$p_z$ plane.}
\label{fig:berry-curvature-purely-spacelike-b}
\end{figure*}

The Berry connections can be written in compact form
using the modified dispersions~\eqref{eq:dispersions-spacelike-b},
\begin{subequations}
\label{eq:berry-connection-spacelike-b}
\begin{align}
\mathcal{A}^{(1,2)}_u&=\frac{1}{2E_u^{(1,2)}(E_u^{(1,2)}\mp p_z+b_z)}\begin{pmatrix}
-p_y \\
p_x \\
0 \\
\end{pmatrix}\,, \displaybreak[0]\\[1ex]
\mathcal{A}^{(1,2)}_v&=\frac{1}{2E_v^{(1,2)}(E_v^{(1,2)}\mp p_z-b_z)}\begin{pmatrix}
-p_y \\
p_x \\
0 \\
\end{pmatrix}\,.
\end{align}
\end{subequations}
The latter are analogous to the standard expressions of Eq.~\eqref{eq:berry-connection-u} when $p_z$ is shifted by $\mp b_z$, where the Weyl nodes are situated. 
The curvatures are found from Eq.~\eqref{eq:berry-curvature-2} to be 
\begin{subequations}
\label{eq:berry-curvature-purely-spacelike-b}
\begin{align}
\Omega_{u,i}^{(1,2)}&=\mp\frac{p_i\mp b_i}{2(E^{(1,2)}_u)^3}\,, \displaybreak[0]\\[1ex]
\Omega_{v,i}^{(1,2)}&=\mp\frac{p_i\pm b_i}{2(E_v^{(1,2)})^3}\,.
\end{align}
\end{subequations}
As in Eq.~\eqref{eq:berry-curvature-isotropic-b}, these are magnetic-monopole analogues, but now shifted by $\pm b_z$ along the $p_z$ axis. 
See Fig.~\ref{fig:magnetic-monopoles-purely-spacelike},
where sink and source monopoles are represented by black dots.
Note that significant Berry flux penetrates the plane through the origin 
and orthogonal to the line connecting the monopoles,
as is visible in Fig.~\ref{fig:berry-curvature-b-contour-plot-x-y}.

Integrating over a 2-sphere $S^2_{0}$ around the origin of momentum space
yields
\begin{align}
\label{eq:berry-curvature-spacelike-case-1}
\int_{S^2_{0}}\mathrm{d}\Omega\,p^2\Omega^{(1,2)}_u\cdot \mathcal{n}&=\int_{S^2_{0}}\mathrm{d}\Omega\,p^2\Omega^{(1,2)}_v\cdot\mathcal{n} \notag \\
&=\mp 2\pi\Theta(p-b_z)\,.
\end{align}
The Heaviside step function $\Theta(x)$ ensures that nonzero contributions only arise when a singularity in the curvature is properly enclosed by the sphere. Analogously, integrations over spheres directly centered at the singularities provide
\begin{align}
\label{eq:berry-curvature-spacelike-case-2}
\int_{S^2_{\mp\mathcal{b}}}\mathrm{d}\Omega\,p^2\Omega^{(1,2)}_u\cdot\mathcal{n}=\mp 2\pi=\int_{S^2_{\pm \mathcal{b}}}\mathrm{d}\Omega\,p^2\Omega^{(1,2)}_v\cdot\mathcal{n}\,.
\end{align}
Contrary to the isotropic case of Sec.~\ref{sec:purely-timelike-sector}, the presence of the preferred direction $\mathcal{b}$ in momentum space separates source and sink of the curvature from each other; see Fig.~\ref{fig:berry-curvature-b-contour-plot-y-z}. The behavior of the curvature along the line separating the monopoles bears similarities to the charge distribution of an electric dipole, where the distribution is constricted in the plane through the origin and orthogonal to the line.

We see that nonzero first Chern numbers 
\begin{equation}
N_u^{(1,2)} =N_v^{(1,2)} =\mp 1
\quad {\rm (purely~spacelike~case)} \,
\label{spacelikechern}
\end{equation}
are associated with any surface that encloses the points $\mathcal{p}=\pm\mathcal{b}$ where the Weyl nodes are situated. This indicates that the Weyl nodes are topologically protected and hence remain when a small fermion mass is introduced. In contrast, the curvature is insensitive to the degeneracies of the two Weyl cones at $p_z=0$, indicated by grey lines in Fig.~\ref{fig:dispersions-purely-spacelike-total}. Indeed, these degeneracies are not topologically protected and are lifted when a small fermion mass emerges.

Let $p_{\pm,i}:=p_i\mp b_i$ be shifted momenta such that $p_{\pm,i}=0$ at the Weyl nodes $p_i=\pm b_i$. 
We can adopt spherical coordinates $(p_{\pm},\vartheta_{\pm},\varphi_{\pm})$ with $p_{\pm}=|\mathcal{p}_{\pm}|$ such that $p_{\pm,x}=p_x=p_{\pm}\sin\vartheta_{\pm}\cos\varphi_{\pm}$, $p_{\pm,y}=p_y=p_{\pm}\sin\vartheta_{\pm}\sin\varphi_{\pm}$, and $p_{\pm,z}=p_z\mp b_z=p_{\pm}\cos\vartheta_{\pm}$. Then, the energy eigenstates of Eq.~\eqref{eq:energy-eigenstates-purely-spacelike-case} satisfy
\begin{subequations}
\begin{align}
\label{eq:eigenstate-u-1}
|u^{(1)}(p_{+,i}+b_i)\rangle&=\frac{\exp(-\mathrm{i}\phi)}{\sqrt{2}}\begin{pmatrix}
\xi^{(1)}(\vartheta_+,\varphi_+) \\
-\xi^{(1)}(\vartheta_+,\varphi_+) \\
\end{pmatrix}\,, \displaybreak[0]\\[1ex]
\label{eq:eigenstate-v-2}
|v^{(2)}(p_{+,i}+b_i)\rangle&=\frac{1}{\sqrt{2}}\begin{pmatrix}
\xi^{(2)}(\vartheta_+,\varphi_+) \\
\xi^{(2)}(\vartheta_+,\varphi_+) \\
\end{pmatrix}\,, \displaybreak[0]\\[1ex]
\label{eq:eigenstate-u-2}
|u^{(2)}(p_{-,i} -b_i)\rangle&=\frac{1}{\sqrt{2}}\begin{pmatrix}
\xi^{(2)}(\vartheta_-,\varphi_-) \\
\xi^{(2)}(\vartheta_-,\varphi_-) \\
\end{pmatrix}\,, \displaybreak[0]\\[1ex]
\label{eq:eigenstate-v-1}
|v^{(1)}(p_{-,i} -b_i)\rangle&=\frac{\exp(-\mathrm{i}\phi)}{\sqrt{2}}\begin{pmatrix}
\xi^{(1)}(\vartheta_-,\varphi_-) \\
-\xi^{(1)}(\vartheta_-,\varphi_-) \\
\end{pmatrix}\,,
\end{align}
\end{subequations}
with the two-component auxiliary spinors of Eq.~\eqref{eq:auxiliary-spinors} evaluated at the new angles $\vartheta_{\pm}$ and $\varphi_{\pm}$.

We can now calculate the momentum-space geometric phases by integrating along circles $S^1_{\pm b_i}$ in the $p_x$-$p_y$ plane centered at $p_i=\pm b_i$. The procedure of the purely timelike case is applicable with the Weyl nodes sitting at $p_{\pm,i}=0$. In terms of the original momentum, the Dirac strings of the eigenstates $|u^{(1)}\rangle$ and $|u^{(2)}\rangle$ are located at $p_z\in (b_z,\infty)$ and $p_z\in (-\infty,-b_z)$, respectively, where $p_x=p_y=0$. The Dirac strings of $|v^{(1)}\rangle$ and $|v^{(2)}\rangle$ run along $p_z\in (-b_z,\infty)$ and $p_z\in (-\infty,b_z)$, respectively, with $p_x=p_y=0$. Therefore, the values of the momentum-space geometric phases are 
\begin{equation}
\Phi^{(1)}_u=-\pi=\Phi^{(1)}_v\,,\quad \Phi^{(2)}_u=\pi=\Phi^{(2)}_v\,.
\label{spacelikephase}
\end{equation}

We can also compute the momentum-space geometric phase around the apices of the Weyl cones in the $p_y$-$p_z$ plane, i.e., for $\varphi_+=\varphi_-=\pi/2$.
The Weyl cones are now separated along $p_z$ in momentum space, an essential feature of this sector. As before, we apply a properly chosen {\rm U(1)} gauge transformation to the eigenstates. It is convenient to adapt Eq.~\eqref{eq:gauge-transformation-dirac-string} to the current setting, where the latter holds in the $p_{\pm,i}$ coordinates with the original Weyl nodes at $\mp b_z$ moved to $p_{\pm,i}=0$. Expressing the spinors and the gauge transformation via the original momentum components $p_i$ yields 
\begin{subequations}
\begin{equation}
|u^{(s)}(\mathcal{p})\rangle\mapsto |u^{(s)}(\mathcal{p})\rangle'=\exp(\mathrm{i}\chi^{(s)})|u^{(s)}(\mathcal{p})\rangle\,,
\end{equation}
where the phases $\chi^{(s)}$ are given by
\begin{align}
\chi^{(1)}&=-\arctan\left[\cot(\varphi_-)-\tan\left(\frac{\vartheta_-}{2}\right)\csc(\varphi_-)\right]\,, \displaybreak[0]\\[1ex]
\chi^{(2)}&=\varphi_++\arctan\left[\cot(\varphi_+)+\tan\left(\frac{\vartheta_+}{2}\right)\csc(\varphi_+)\right]\,.
\end{align}
\end{subequations}
In an analogous manner, 
results for the $v$-type eigenstates are found. Here, the angles $\vartheta_{\pm}$ and $\varphi_{\pm}$ must now be expressed in terms of $p_i$ and $b_z$ according to their definitions. As a result, the Dirac strings are relocated to straight lines parallel to the $p_x$ axis. The Dirac strings for $|u^{(1)}\rangle'$, $|v^{(1)}\rangle'$ run along the sets $p_x\in (0,\infty)$, $p_y=0$, and $p_z=\pm b_z$ and for $|u^{(2)}\rangle'$, $|v^{(2)}\rangle'$ we have $p_x\in (-\infty,0)$, $p_y=0$, and $p_z=\mp b_z$. This enables us to integrate along the circles in Fig.~\eqref{fig:dispersions-purely-spacelike-total} in the $p_y$-$p_z$ plane, which reproduces the momentum-space geometric phases of Eq.~\eqref{spacelikephase}.

\subsection{Generic spacelike coefficient}

As a generalization of the purely spacelike sector, we consider the case of a spacelike $b_{\mu}$ with nonzero $b_0$ and $b_z$,
with hamiltonian 
\begin{equation}
H_b=H_0-b_0\gamma_5+b_z\gamma^0\gamma_5\gamma^z\,,
\end{equation}
and the reinterpreted dispersions
\begin{subequations}
\label{eq:dispersions-spacelike}
\begin{align}
E_u^{(1,2)}&=|\mathcal{p}\mp \mathcal{b}|\pm b_0\,, \\[1ex]
E_v^{(1,2)}&=|\mathcal{p}\pm \mathcal{b}|\mp b_0\,.
\end{align}
\end{subequations}
Now, the Weyl cones are shifted along the energy axis by $\pm b_0$ and along the $p_z$ axis by $\pm b_z$. Both dispersions were discussed in detail in Ref.~\cite{AlanKostelecky:2024gek} and our solution of the concordance problem applies to the latter scenario. Note that $[H_b,H_{\mathcal{b}}]=0$ with $H_{\mathcal{b}}$ of Eq.~\eqref{eq:hamiltonian-purely-spacelike} for the purely spacelike case. Although the dispersions are different, the energy eigenstates for the purely spacelike sector quoted in Eq.~\eqref{eq:energy-eigenstates-purely-spacelike-case} can be taken over without any modifications. Therefore, the curvatures~\eqref{eq:berry-curvature-purely-spacelike-b} as well as the first Chern numbers~\eqref{spacelikechern} still hold.

\begin{figure}
\centering
\subfloat[]{\label{fig:dispersions-spacelike}\includegraphics[scale=0.2]{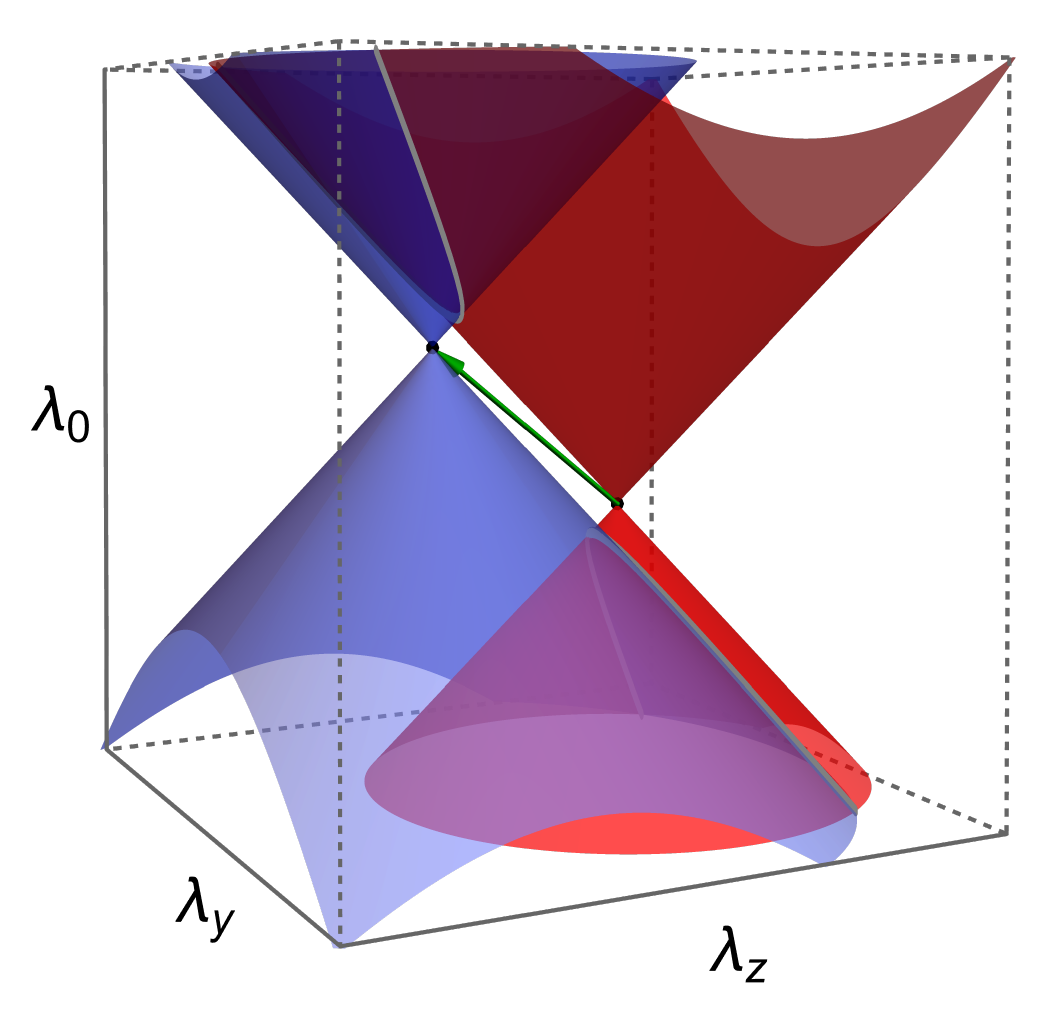}}
\subfloat[]{\label{fig:dispersions-spacelike-reinterpreted}\includegraphics[scale=0.2]{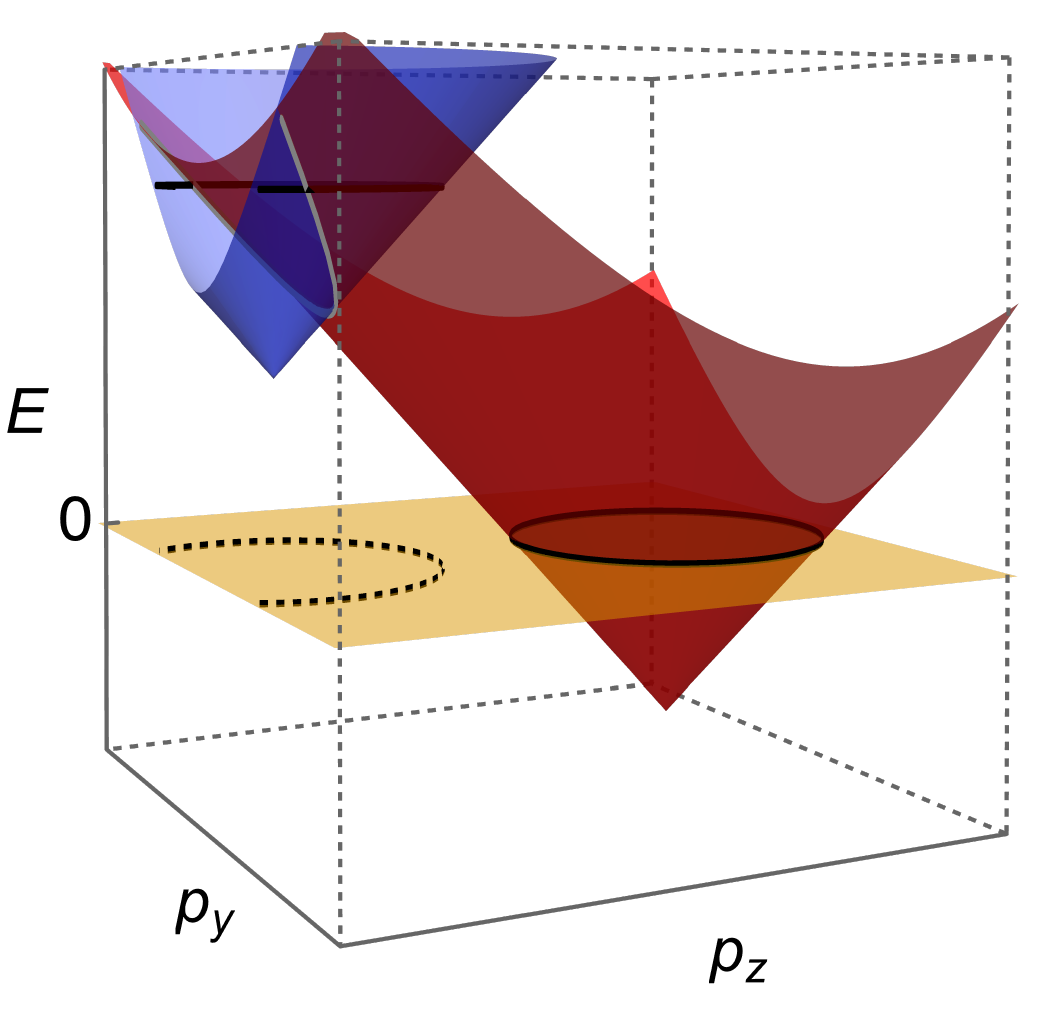}}
\caption{Energy eigenvalues for the generic spacelike case \protect\subref{fig:dispersions-spacelike} before and \protect\subref{fig:dispersions-spacelike-reinterpreted} after reinterpretation.}
\label{fig:dispersions-spacelike-total}
\end{figure}

\begin{figure}
\centering
\includegraphics[scale=0.3]{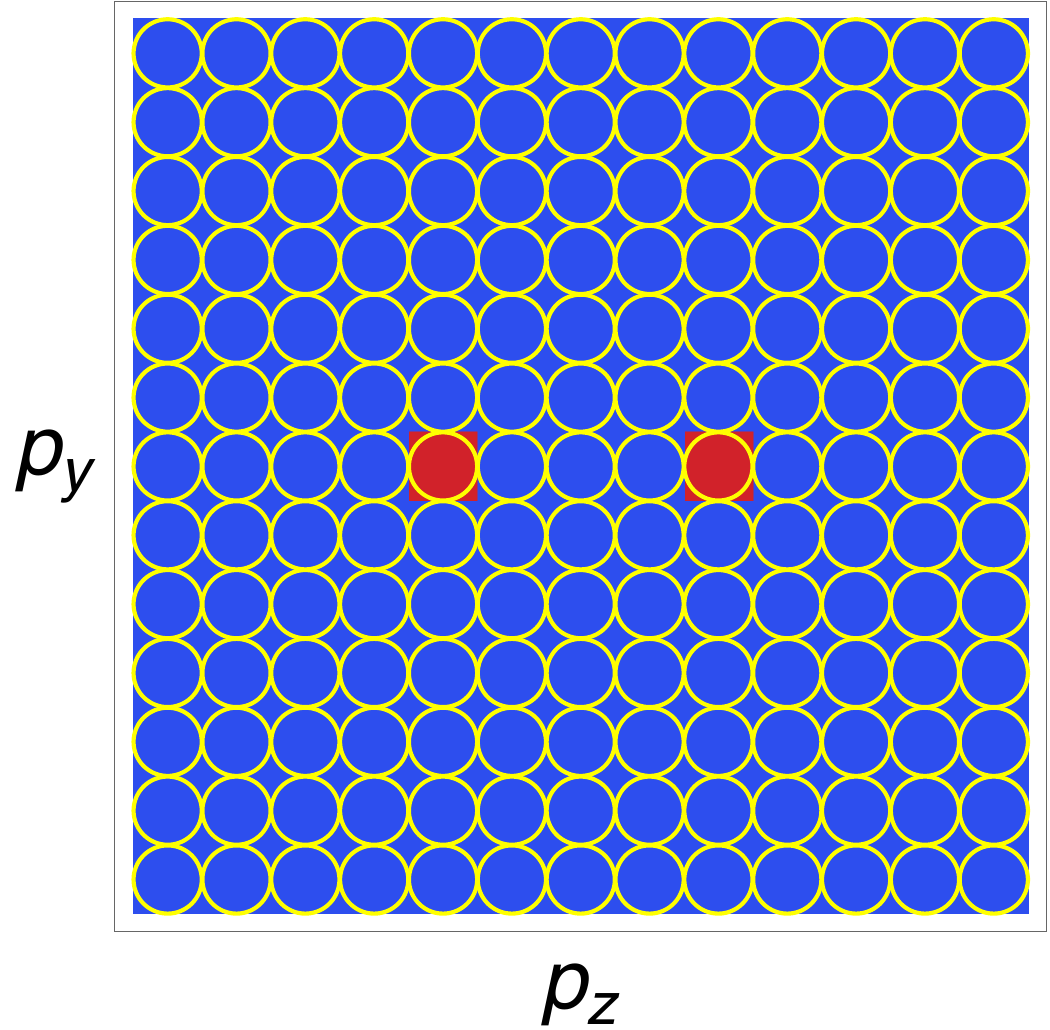}
\caption{Mesh of circles in the $p_y$-$p_z$ plane.}
\label{fig:berry-phases-covering-plane}
\end{figure}

In our recent paper \cite{AlanKostelecky:2024gek} we were interested in the momentum-space geometric phases for the $u$- and $v$-type dispersions that intersect the Fermi plane. These are $E_u^{(2)}$ and $E_v^{(1)}$, represented by the red cone in Fig.~\ref{fig:dispersions-spacelike-reinterpreted}. As before, we integrate around circles $S^1_{\pm\mathcal{b}}$ lying in the $p_x$-$p_z$ plane and encircling the apices of the corresponding Weyl cones. The shift of the cones along the energy axis caused by $b_0$ turns out to have no impact on the phases. By following the procedure of Sec.~\ref{sec:purely-spacelike-b}, we find
\begin{equation}
\Phi'^{(2)}_u=\pi\,,\quad \Phi'^{(1)}_v=-\pi\,.
\end{equation}
This result can be verified numerically by
integrating along small circles centered at the sites of a lattice 
covering the plane, 
as illustrated in Fig.~\ref{fig:berry-phases-covering-plane}. 
Nonzero contributions of $\Phi=\pm\pi$ arise 
for the circles on the red squares in the figure, 
which enclose the apices of the Weyl cones in $\mathcal{P}_3$.
In contrast, any circle without an apex inside contributes $\Phi=0$. This method provides a numerical alternative for detecting the presence of Weyl nodes via the Berry connection.

The ground state of the theory with a generic spacelike $b_\mu$
has recently been studied using thermodynamic techniques
and shown to be populated with fermions and antifermions
of negative energies~\cite{AlanKostelecky:2024gek},
unlike the vacuum in a Lorentz-invariant setting
where no particles are present.
This nontrivial ground state resembles that of electrons and holes 
in a one-particle description of certain Weyl semimetals~\cite{Lv:2021oam}. 
The fermions and antifermions in the vacuum  
can be excited to positive-energy states
via the exchange of energy with the thermodynamic bath.
An excitation can acquire a momentum-space geometric phase 
if it traverses a loop enclosing the apex of the corresponding Weyl cone, 
as indicated in Fig.~\ref{fig:dispersions-spacelike-reinterpreted}.
The excitation can then drop back to the vacuum
by releasing energy to the bath.
The vacuum state itself therefore can acquire
a nontrivial momentum-space geometric phase.

In the limit as the temperature $T$ of the bath is taken to zero,
the phase of the ground state freezes to a particular value $\Phi_0$.
Since individual contributions to $\Phi_0$
from each fermion or antifermion are quantized,
$\Phi_0$ itself is quantized.
Moreover,
the phase is topologically protected in this limit.
We therefore can conclude that the vacuum 
in the theory with a generic spacelike $b_\mu$
can be chosen from among an infinite number of degenerate ground states,
each characterized by a momentum-space geometric phase
that is also topological.
Investigating the possible phenomenological implications of this feature
would be of definite interest
but lies beyond our present scope.

\section{CPT-even case}
\label{sec:application-sme-d}

In this section,
the momentum-space geometric phases 
associated with the action $S_d$ of Eq.~\eqref{dact} are investigated.
We examine in turn 
cases with isotropic, mixed, and spacelike $d_{\mu\nu}$ coefficients. 
The first Chern numbers are derived in each case,
and the corresponding momentum-space geometric phases
are then discussed.

\subsection{Isotropic sector}
\label{sec:isotropic-d}

The tracelessness of $d_{\mu\nu}$
implies a single isotropic combination of components exists.
This isotropic combination can be written in diagonal matrix form as
\begin{equation}
d_{\mu\nu}=d_{00}
\begin{pmatrix}
1 & 0 & 0 & 0 \\
0 & \tfrac 13 & 0 & 0 \\
0 & 0 & \tfrac 13 & 0 \\
0 & 0 & 0 & \tfrac 13 \\
\end{pmatrix}\,.
\end{equation}
The corresponding Lagrange density involves an additional time derivative, which leads to an unconventional time evolution of asymptotic states. To restore conventional time evolution, the responsible terms can be removed from the Lagrange density via a suitable field redefinition. We follow the method presented in Ref.~\cite{Colladay:2001wk} and search for an invertible matrix $A$ that satisfies $A^{\dagger}\gamma^0\Gamma^0A=\mathds{1}_4$ with $\Gamma^0=\gamma^0+d^{\mu0}\gamma_5\gamma_{\mu}$, where $\mathds{1}_n$ is the $(n\times n)$ identity matrix. 

A possible choice for the matrix $A$ is
\begin{subequations}
\label{eq:inverse-A-isotropic-d}
\begin{align}
A&=\left(\begin{array}{c|c}
\mathcal{C} & \mathcal{B} \\
\hline
\mathcal{B} & \mathcal{C} \\
\end{array}\right)\,,\quad \mathcal{C}=A^+\mathds{1}_2\,,\quad \mathcal{B}=A^-\mathds{1}_2\,, \\[1ex]
A^{\pm}&=\sqrt{\frac{1\pm \sqrt{1-(d_{00})^2}}{2(1-(d_{00})^2)}}\,,
\end{align}
\end{subequations}
with inverse 
\begin{equation}
\label{eq:inserve-A}
A^{-1}=\sqrt{1-(d_{00})^2}\left(\begin{array}{c|c}
\mathcal{C} & -\mathcal{B} \\
\hline
-\mathcal{B} & \mathcal{C} \\
\end{array}\right)\,,
\end{equation}
Note that this transformation is only valid for $d_{00}\in (-1,1)$, as $A$ becomes either singular or complex otherwise. 

Let $\psi$ be the Dirac field in the original SME action $S_d$.
Then, a new Dirac field $\chi:=A^{-1}\psi$ is introduced such that the latter satisfies a time-dependent Schr\"{o}dinger-type equation $H_{\ring{d}}\,\chi=\mathrm{i}\partial_0\chi$ with Hermitian hamiltonian
\begin{equation}
\label{eq:transformed-hamiltonian-isotropic}
H_{\ring{d}}=\gamma^0\overline{A}(\gamma^j+d^{\mu j}\gamma_5\gamma_{\mu})\lambda^jA\,,\quad \overline{A}:=\gamma^0A^{\dagger}\gamma^0\,,
\end{equation}
where $\ring{d}$ denotes the isotropic piece of $d_{\mu\nu}$~\cite{Kostelecky:2013rta}. The time evolution of asymptotic states induced by the new hamiltonian is then conventional. The dispersions remain unaltered by the field redefinition
\begin{subequations}
\label{eq:dispersion-relations-isotropic-d}
\begin{align}
E_u^{(1,2)}&=\frac{1\mp d_{00}/3}{|1\pm d_{00}|}p\,, \\[1ex]
E_v^{(1,2)}&=\frac{1\mp d_{00}/3}{|1\pm d_{00}|}p\,.
\end{align}
\end{subequations}
They describe two distinct Weyl cones with decreased and increased opening angle, respectively, as compared to the Lorentz-invariant case. See Figs.~\ref{fig:dispersions-d-isotropic} and \ref{fig:dispersions-d-isotropic-reinterpreted} for 
the case $d_{00}=1/3$. Note that norms are computed only in the denominators. The dispersions have singularities at $d_{00}=\pm 1$, which are already excluded from the interval of validity of Eq.~\eqref{eq:inverse-A-isotropic-d}. Negative energies, which would occur for $d_{00}>3$ or $d_{00}\leq -3$, are thereby also avoided.

The standard Dirac hamiltonian of Eq.~\eqref{eq:dirac-hamiltonian-standard} commutes with the hamiltonian $H_{\ring{d}}$ of Eq.~\eqref{eq:transformed-hamiltonian-isotropic} redefined by $A$. Hence, the energy eigenstates after the field redefinition are the standard ones of Eq.~\eqref{eq:energy-eigenstates-timelike}.
Then, the Berry connection and curvature are unaffected by Lorentz violation, i.e., Eqs.~\eqref{eq:berry-connection-u} and \eqref{eq:berry-curvature-isotropic-b} still hold. 
This implies the first Chern numbers 
\begin{equation}
N_u^{(1,2)}= N_v^{(1,2)}= \mp 1 
\quad {\rm (isotropic~case)} \,.
\end{equation}

\subsection{Mixed sector $\boldsymbol{d_{i0}}$}
\label{sec:mixed-d-1}

Next, we explore the mixed components $d_{i0}$ with all others set to zero. Since $d_{i0}$ are accompanied by an additional time derivative in the Lagrange density, we apply the method used in the isotropic case. The Dirac spinor field is to be transformed via a matrix $A$ that obeys $A^{\dagger}\gamma^0\Gamma^0A=\mathds{1}_4$ with $\Gamma^0=\gamma^0+d^{i0}\gamma_5\gamma_i$. 
Via a generic \textit{ansatz} based on the complete set of 16 matrices in spinor space, we can demonstrate that a suitable matrix $A$ can be given by 
\begin{subequations}
\begin{align}
A&=\left(\begin{array}{c|c}
\mathds{1}_2-\mathcal{D} & \mathcal{E} \\
\hline
\mathcal{E} & \mathds{1}_2-\mathcal{D} \\
\end{array}
\right)\,, \displaybreak[0]\\[1ex]
\mathcal{D}&=-\frac{\mathcal{d}\cdot\boldsymbol{\sigma}}{2(1-\mathcal{d}^2)}\,,\quad
\mathcal{E}=\sqrt{3-4\mathcal{d}^2}\mathrm{i}\mathcal{D}\,,
\end{align}
\end{subequations}
with inverse 
\begin{subequations}
\label{eq:inverse-matrix-A-mixed}
\begin{align}
A^{-1}&=\frac{1}{1-\mathcal{d}^2}\left(\begin{array}{c|c}
\tilde{\mathcal{D}} & \tilde{\mathcal{E}} \\
\hline
\tilde{\mathcal{E}} & \tilde{\mathcal{D}} \\
\end{array}\right)\,, \displaybreak[0]\\[1ex]
\tilde{\mathcal{D}}&=\left(1-\frac{3}{2}\mathcal{d}^2\right)\mathds{1}_2-\frac{1}{2}(1-2\mathcal{d}^2)\mathcal{d}\cdot\boldsymbol{\sigma}\,, \displaybreak[0]\\[1ex]
\tilde{\mathcal{E}}&=-\frac{\mathrm{i}}{2}\sqrt{3-4\mathcal{d}^2}(\mathcal{d}^2\mathds{1}_2-\mathcal{d}\cdot\boldsymbol{\sigma})\,,
\end{align}
\end{subequations}
conveniently expressed in terms of the Pauli matrices $\boldsymbol{\sigma}=(\sigma^x,\sigma^y,\sigma^z)$ and the form $\mathcal{d}:=(d_{i0})$. The dispersions are readily obtained as
\begin{subequations}
\label{eq:dispersion-relations-mixed-d-1}
\begin{align}
E_u^{(1,2)}&=\frac{\pm \mathcal{d}\cdot\mathcal{p}+\Upsilon}{1-\mathcal{d}^2}=E_v^{(1,2)}\,, \\[1ex]
\Upsilon&=\Upsilon(\mathcal{p})=\sqrt{(1-\mathcal{d}^2)\mathcal{p}^2+(\mathcal{d}\cdot\mathcal{p})^2}\,.
\end{align}
\end{subequations}
Therefore, $\mathcal{d}$ tilts the Weyl cones and the presence of the factor $1-\mathcal{d}^2$ implies changes of their opening angles. Note that both Weyl cones are tilted in opposite directions, whereas the modified opening angle is the same for both cones.
Figures~\ref{fig:dispersions-d-mixed-1} and \ref{fig:dispersions-d-mixed-1-reinterpreted} display these features for $d_{10}=d_{20}=0$ and $d_{30}=1/2$.

As in the isotropic case, a new Dirac spinor $\chi:=A^{-1}\psi$ is defined such that the latter obeys a time-dependent Schr\"{o}dinger-type equation $H_{\mathcal{d}}\,\chi=\mathrm{i}\partial_0\chi$ with Hermitian hamiltonian
\begin{equation}
H_{\mathcal{d}}=\gamma^0\overline{A}\gamma^j\lambda^jA\,,
\end{equation}
with the matrix $\overline{A}$ defined according to Eq.~\eqref{eq:transformed-hamiltonian-isotropic}. Now, the energy eigenstates for modified Dirac fermions are of the form
\begin{subequations}
\begin{align}
|u^{(1,2)}(\mathcal{p})\rangle&=|v^{(1,2)}(\mathcal{p})\rangle \notag \\
&=\mathcal{N}^{(1,2)}A^{-1}
\begin{pmatrix}
\pm (\bar{p}_x-\mathrm{i}\bar{p}_y) \\
\Upsilon \mp \bar{p}_z \\
-(\bar{p}_x-\mathrm{i}\bar{p}_y) \\
\mp \Upsilon+\bar{p}_z \\
\end{pmatrix}\,,
\end{align}
with $\bar{p}_i:=p_i-\mathrm{i}(\mathcal{d}\times\mathcal{p})_i$, the inverse matrix $A^{-1}$ given in Eq.~\eqref{eq:inverse-matrix-A-mixed},
and the normalizations
\begin{align}
\mathcal{N}^{(1,2)}&=\frac{1}{2}\Big(\Upsilon[\mp(1-\mathcal{d}^2)p_z \notag \\
&\phantom{{}={}}\quad+(1-d_{z0})(\Upsilon\mp\mathcal{d}\cdot\mathcal{p})]\Big)^{-\frac{1}{2}}\,.
\end{align}
\end{subequations}
These equations permit us to compute the Berry connections. 
After some algebra, we find
\begin{align}
\label{eq:berry-connection-mixed-d-2}
\mathcal{A}_u^{(1,2)}&=\frac{1-\mathcal{d}^2}{2\Upsilon[(1+d_{z0})(\Upsilon\mp\mathcal{d}\cdot\mathcal{p})\mp(1-\mathcal{d}^2)p_z]} \notag \\
&\phantom{{}={}}\times\begin{pmatrix}
-p_y+(\mathcal{d}\times\mathcal{p})_x \\
p_x+(\mathcal{d}\times\mathcal{p})_y \\
(\mathcal{d}\times\mathcal{p})_z \\
\end{pmatrix}=\mathcal{A}_v^{(1,2)}\,.
\end{align}
The result incorporates a modification dependent on $\mathcal{d}$.
The Dirac strings run along sets of points where the denominators of Eq.~\eqref{eq:berry-connection-mixed-d-2} vanish. These sets are intricate for nonzero $\mathcal{d}$. 

The curvatures follow directly from Eq.~\eqref{eq:berry-connection-mixed-d-2}.
They read
\begin{equation}
\Omega_{u,i}^{(1,2)}=\mp\frac{(1-\mathcal{d}^2)p_i}{2\Upsilon^3(\mathcal{p})}=\Omega_{v,i}^{(1,2)}\,.
\end{equation}
Note that these expressions reduce as required to the standard results~\eqref{eq:berry-connection-u} and \eqref{eq:berry-curvature-isotropic-b}, respectively, 
for $d_{i0}\rightarrow 0$. Note also that the curvatures are unaffected by Lorentz violation at the linear level. Moreover, as in the Lorentz-invariant case, $\Omega_{u,i}^{(1)}+\Omega_{u,i}^{(2)}=0=\Omega_{v,i}^{(1)}+\Omega_{v,i}^{(2)}$. The singularities are situated at $p_i=0$ and the first Chern numbers can be computed in spherical coordinates,
yielding the results 
\begin{equation}
N_u^{(1,2)}= N_v^{(1,2)}= \mp 1
\quad {\rm (first~mixed~case)} \,.
\end{equation}

\subsection{Mixed sector $\boldsymbol{d_{0i}}$}
\label{sec:mixed-d-2}

\begin{figure}[h]
\centering
\subfloat[]{\label{fig:dispersions-d-isotropic}\includegraphics[scale=0.2]{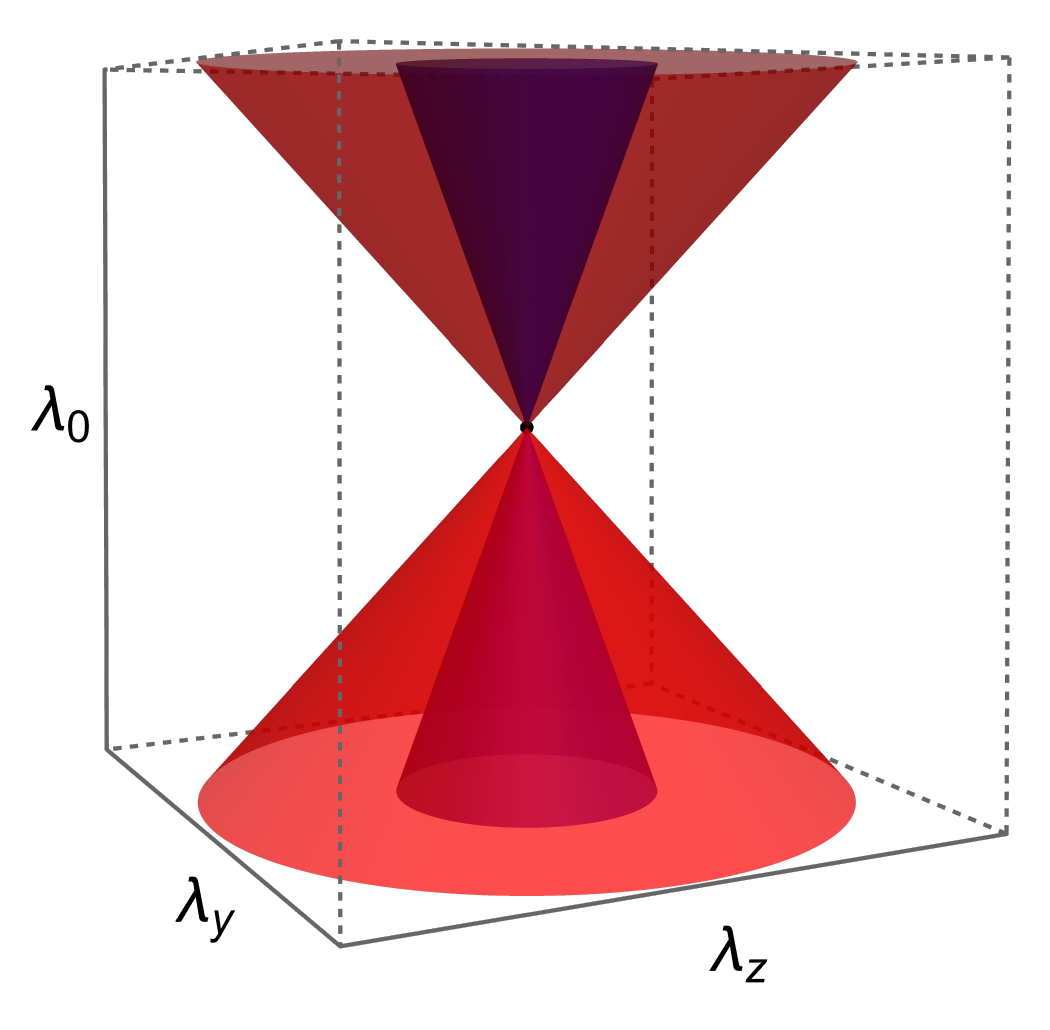}}\hspace{0.4cm}
\subfloat[]{\label{fig:dispersions-d-isotropic-reinterpreted}\includegraphics[scale=0.2]{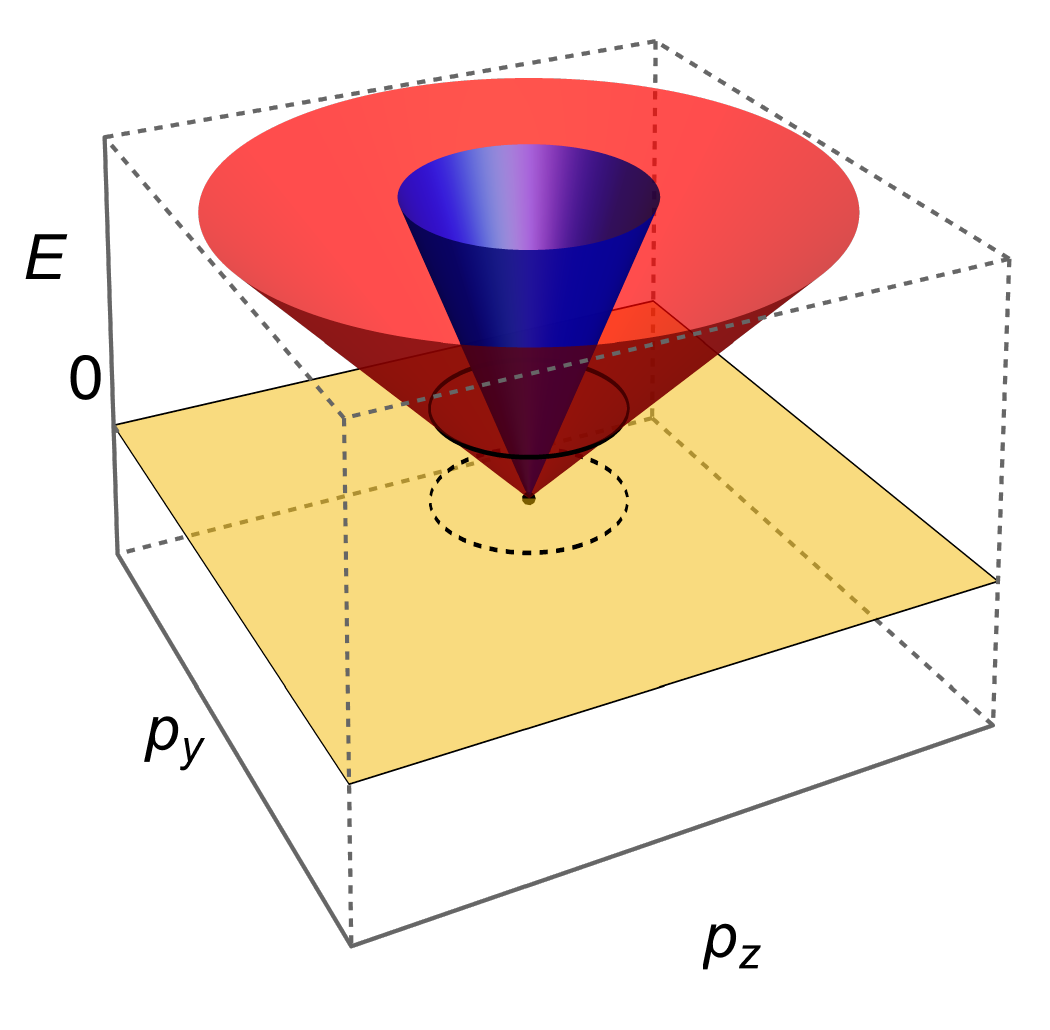}} \\
\subfloat[]{\label{fig:dispersions-d-mixed-1}\includegraphics[scale=0.2]{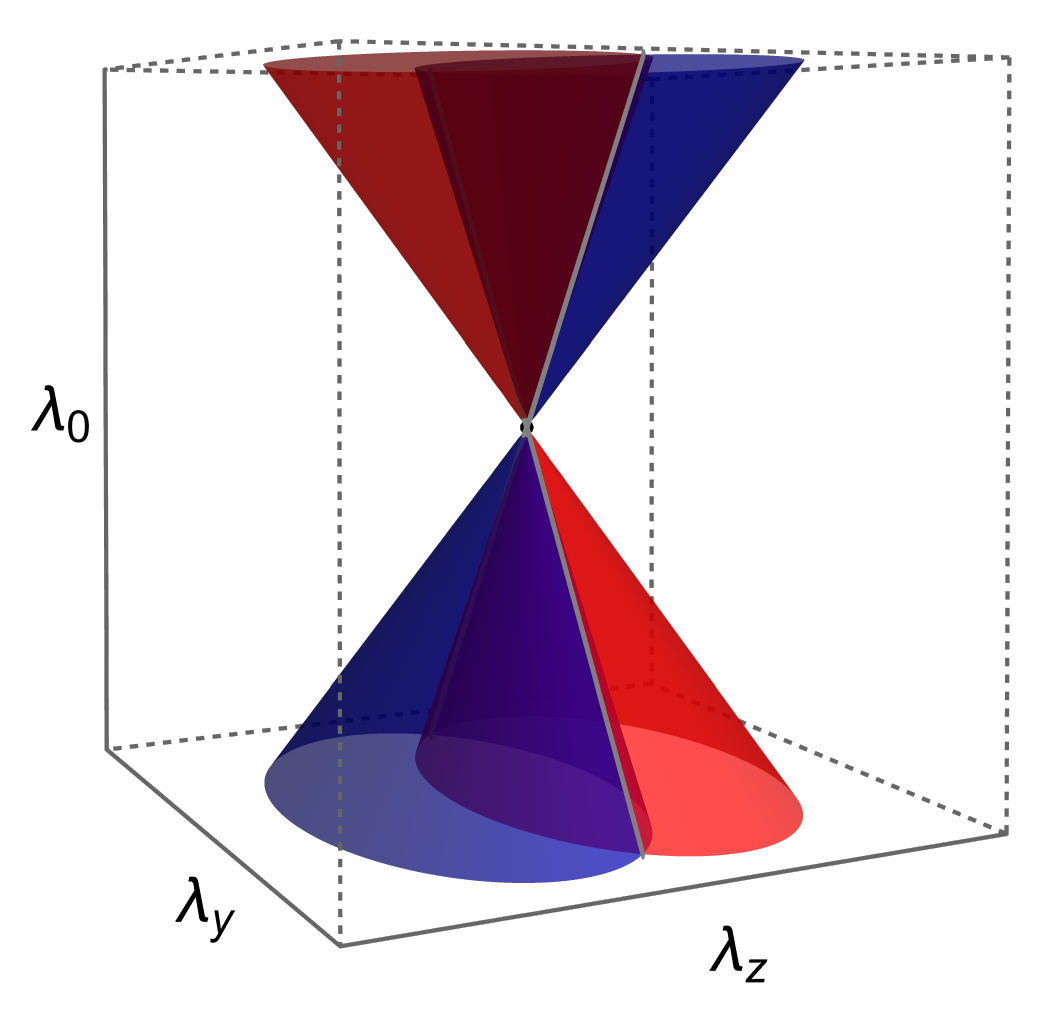}}\hspace{0.4cm}
\subfloat[]{\label{fig:dispersions-d-mixed-1-reinterpreted}\includegraphics[scale=0.2]{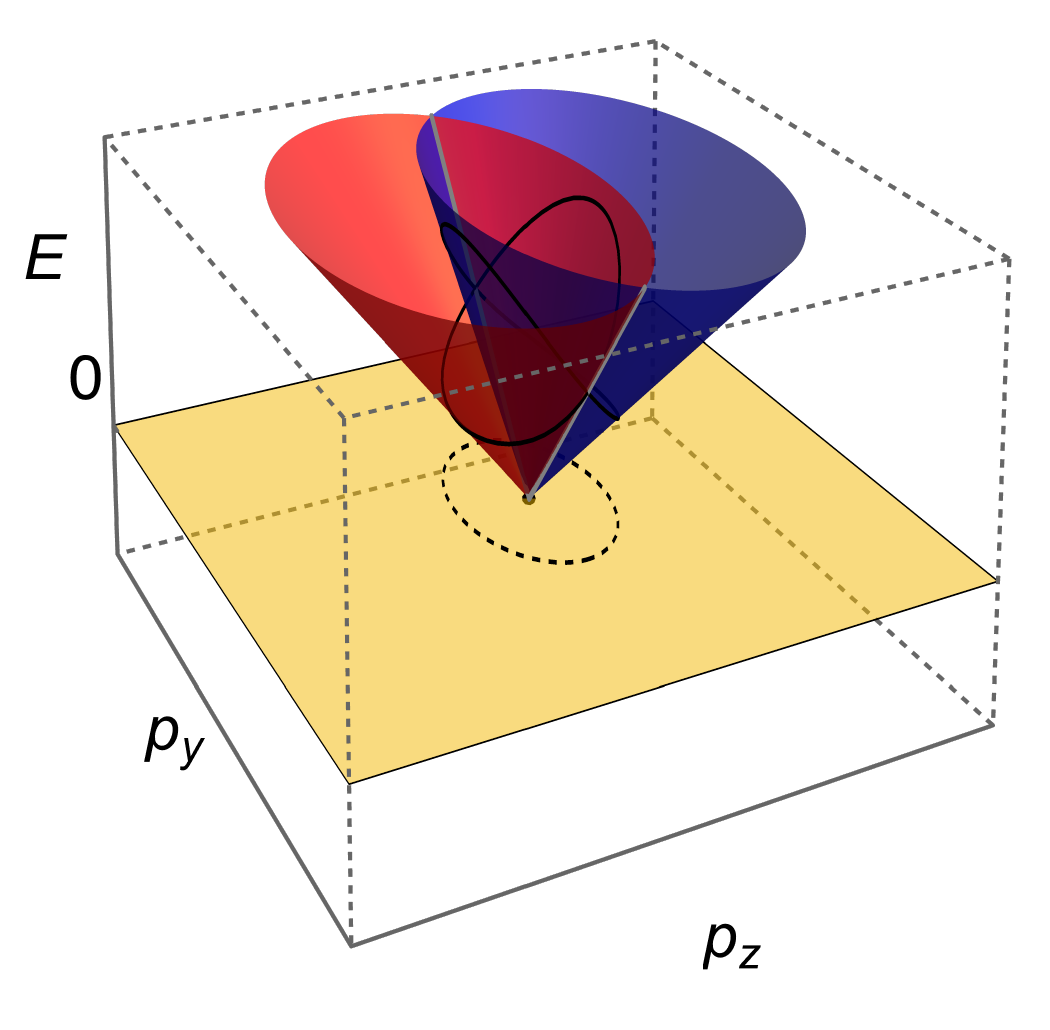}} \\
\subfloat[]{\label{fig:dispersions-d-mixed-2}\includegraphics[scale=0.2]{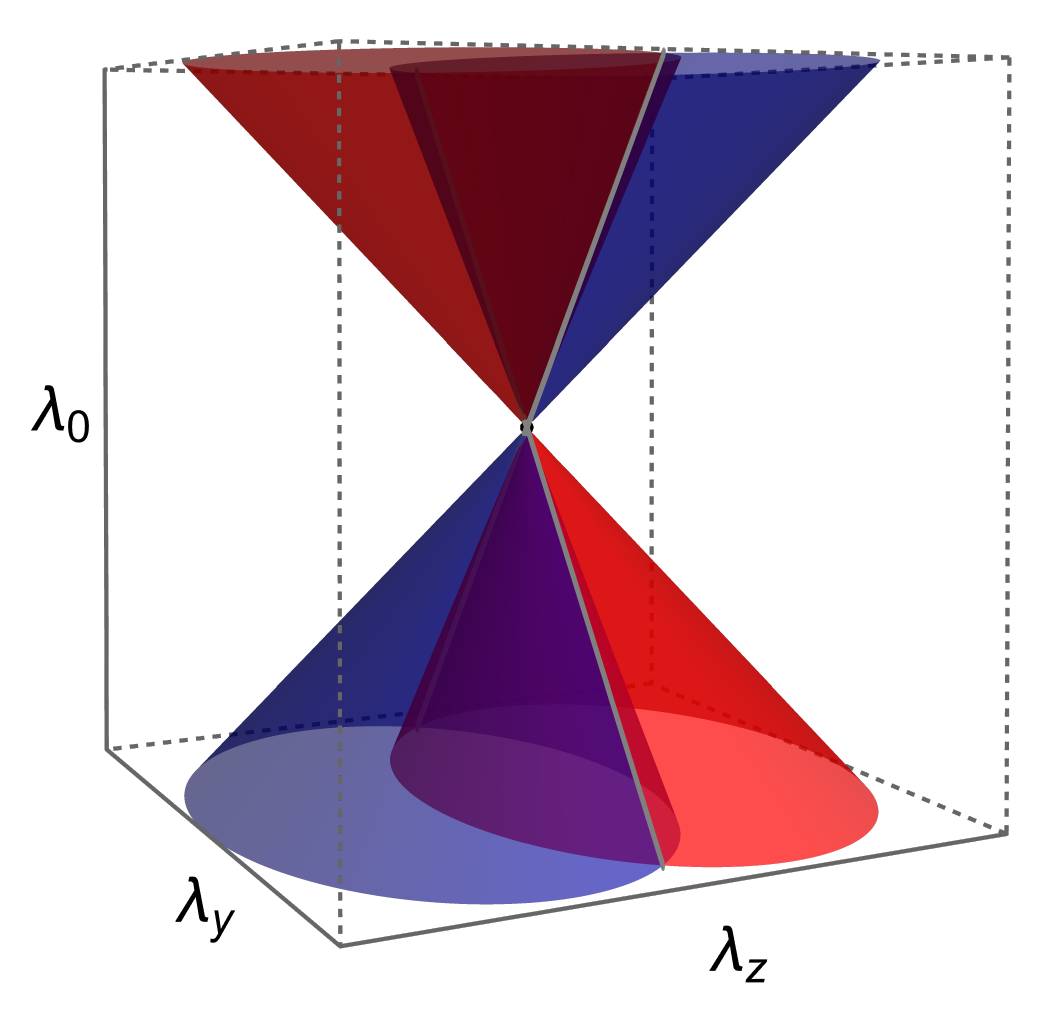}}\hspace{0.4cm}
\subfloat[]{\label{fig:dispersions-d-mixed-2-reinterpreted}\includegraphics[scale=0.2]{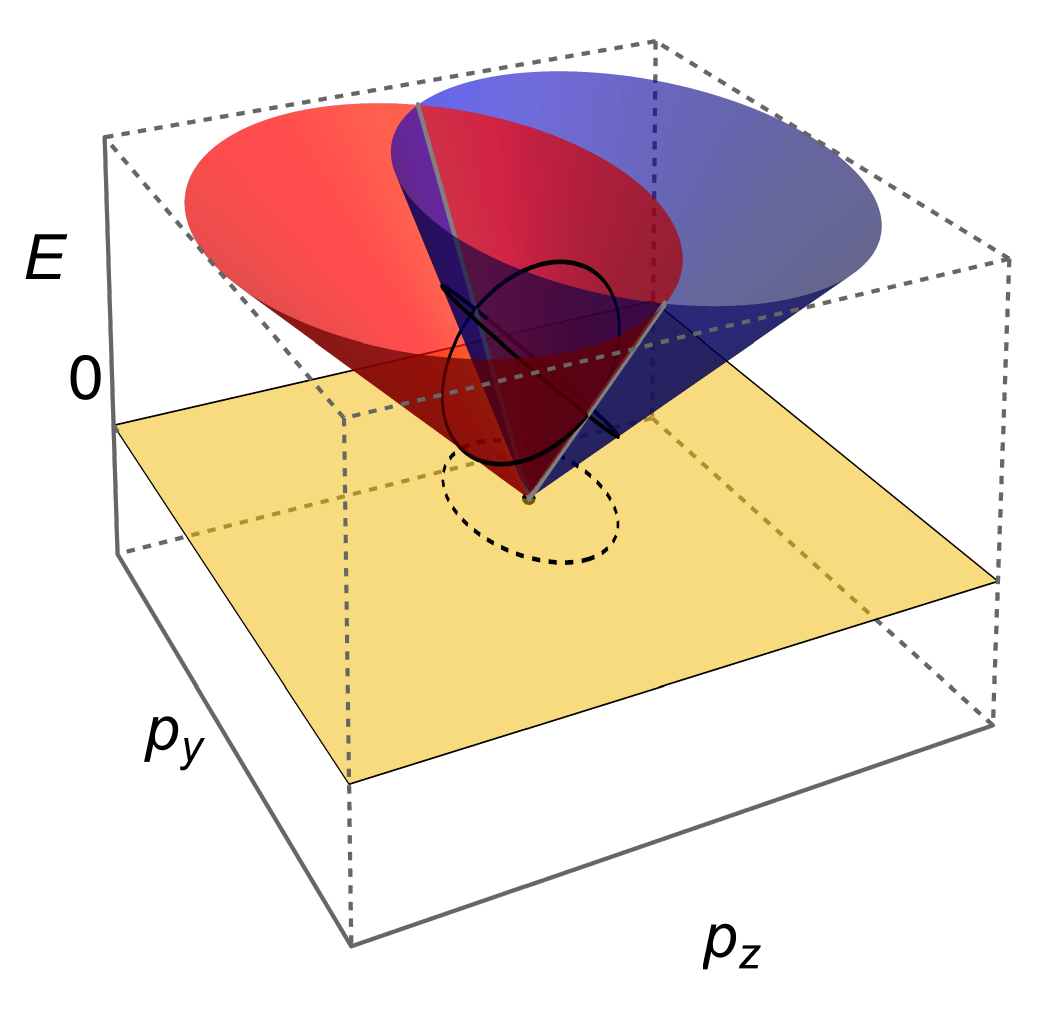}} \\
\subfloat[]{\label{fig:dispersions-d-spacelike}\includegraphics[scale=0.2]{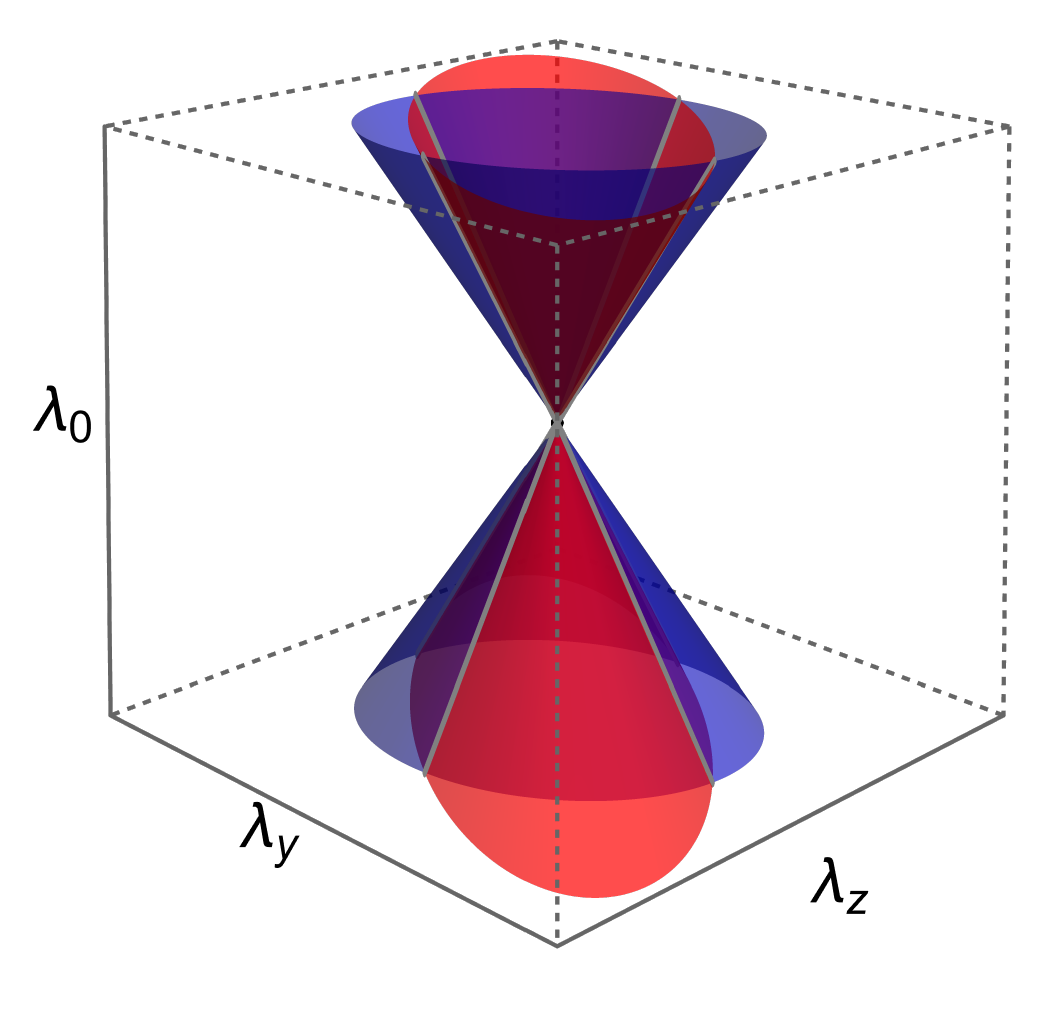}}\hspace{0.4cm}
\subfloat[]{\label{fig:dispersions-d-spacelike-reinterpreted}\includegraphics[scale=0.2]{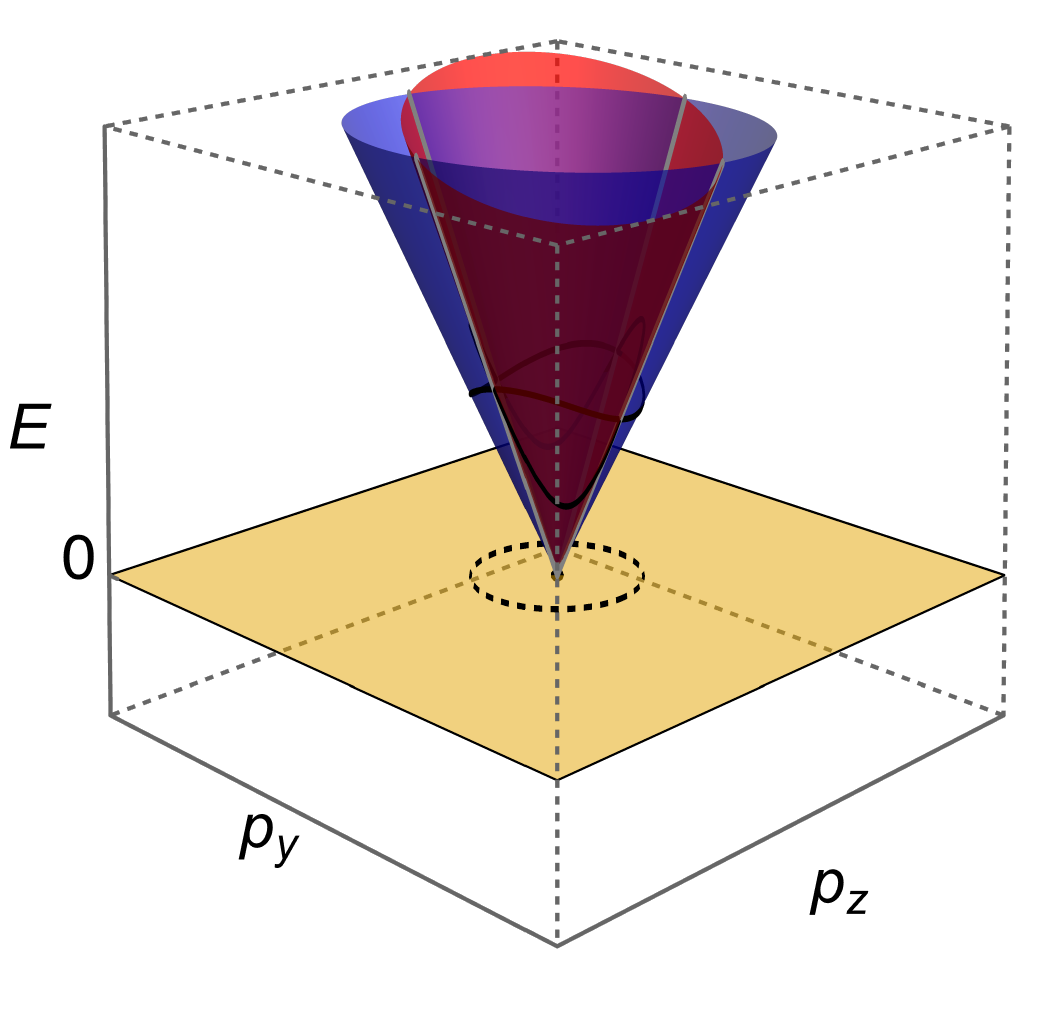}}
\caption{Energy eigenvalues before and after reinterpretation for 
\protect\subref{fig:dispersions-d-isotropic}, 
\protect\subref{fig:dispersions-d-isotropic-reinterpreted}
the isotropic case~\eqref{eq:dispersion-relations-isotropic-d},
\protect\subref{fig:dispersions-d-mixed-1},
\protect\subref{fig:dispersions-d-mixed-1-reinterpreted}
the first mixed case~\eqref{eq:dispersion-relations-mixed-d-1},
\protect\subref{fig:dispersions-d-mixed-2},
\protect\subref{fig:dispersions-d-mixed-2-reinterpreted}
the second mixed case~\eqref{eq:dispersion-relations-mixed-d-2},
and \protect\subref{fig:dispersions-d-spacelike},
\protect\subref{fig:dispersions-d-spacelike-reinterpreted}
the purely spacelike case~\eqref{eq:dispersions-d-spacelike}.
}
\label{fig:dispersions-d-mixed}
\end{figure}

This section concerns the mixed coefficients $d_{0i}$, with all others set to zero. It is helpful to introduce the three-component form $\tilde{\mathcal{d}}:=(d_{0i})$. Unlike the previous isotropic and mixed configurations of Secs.~\ref{sec:isotropic-d} and \ref{sec:mixed-d-1}, respectively, the coefficients $d_{0i}$ are not contracted with an additional time derivative. It is therefore unnecessary to redefine the Dirac spinor field. The hamiltonian reads
\begin{equation}
H_{\tilde{\mathcal{d}}}=H_0+d_{0i}\gamma^0\gamma_5\gamma^0\lambda_i\,,
\end{equation}
where $H_0$ is the Dirac hamiltonian of Eq.~\eqref{eq:dirac-hamiltonian-standard}. The dispersions are
\begin{equation}
\label{eq:dispersion-relations-mixed-d-2}
E_u^{(1,2)}=\pm\tilde{\mathcal{d}}\cdot\mathcal{p}+p=E_v^{(1,2)}\,.
\end{equation}
This reveals that
$\tilde{\mathcal{d}}$ tilts the Weyl cones, while preserving their opening angles. See Figs.~\ref{fig:dispersions-d-mixed-2} and \ref{fig:dispersions-d-mixed-2-reinterpreted} for the case $d_{01}=d_{02}=0$ and $d_{03}=1/2$.

The vanishing commutator $[H_0,H_{\tilde{\mathcal{d}}}]=0$ implies that the eigenstates of the Dirac hamiltonian remain unaffected by the coefficients $d_{0i}$, i.e., they can be taken over from Eq.~\eqref{eq:energy-eigenstates-timelike}. 
The curvature therefore takes the same form as in Eq.~\eqref{eq:berry-curvature-isotropic-b}. The tilting of the Weyl cones has no effect, so the integrated curvatures are given by Eq.~\eqref{eq:berry-connection-integrated}. We thus obtain the first Chern numbers as 
\begin{equation}
N_u^{(1,2)}= N_v^{(1,2)}= \mp 1 
\quad {\rm (second~mixed~case)} \,.
\end{equation}

\subsection{Purely spacelike off-diagonal coefficients $\boldsymbol{d_{ij}}$}
\label{sec:purely-spacelike-d}

\begin{figure*}[t]
\centering
\subfloat[]{\label{fig:berry-curvature-d-contour-plot-x-y}\includegraphics[scale=0.13]{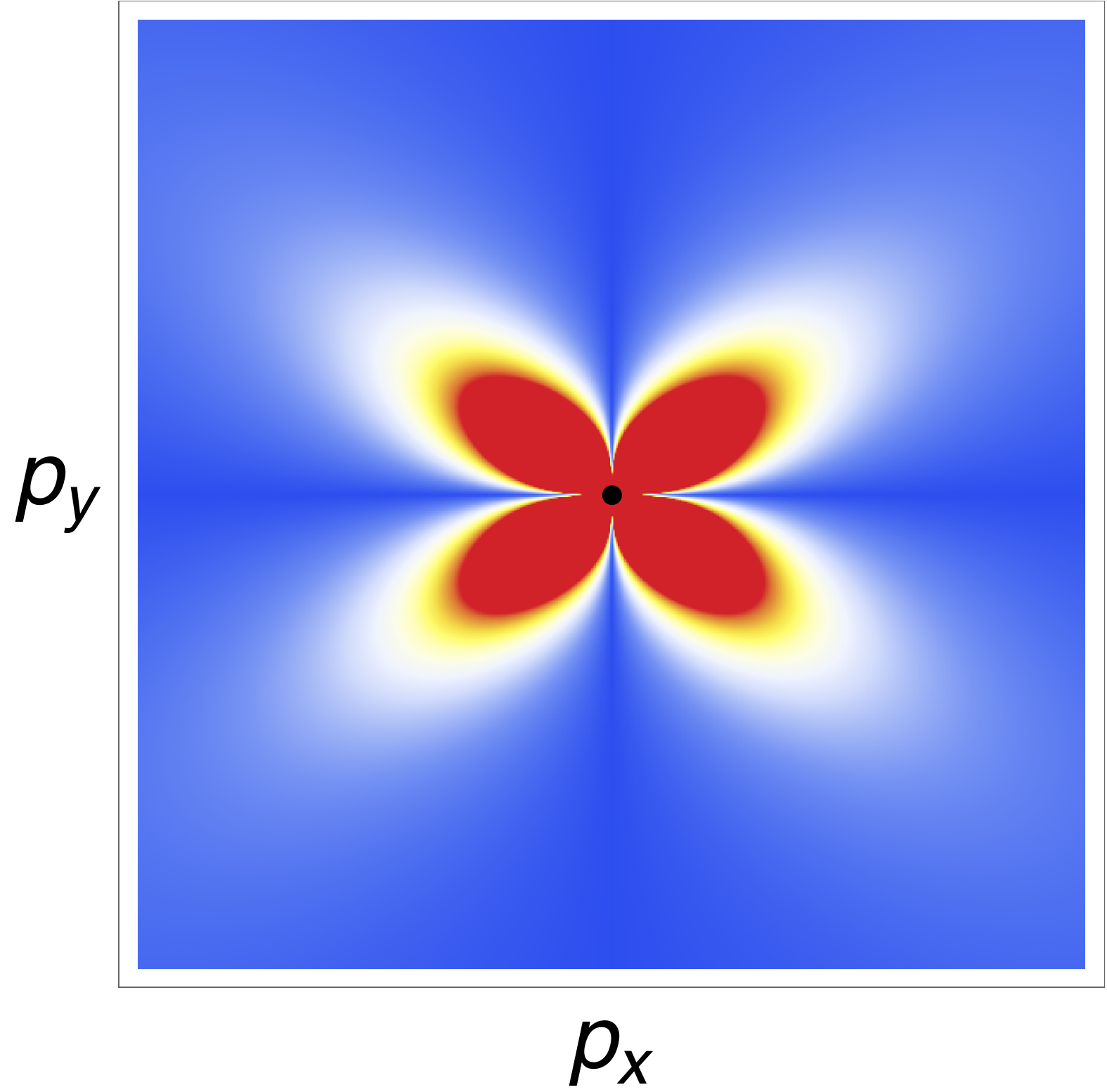}}\hspace{1cm}
\subfloat[]{\label{fig:berry-curvature-d-contour-plot-x-z}\includegraphics[scale=0.13]{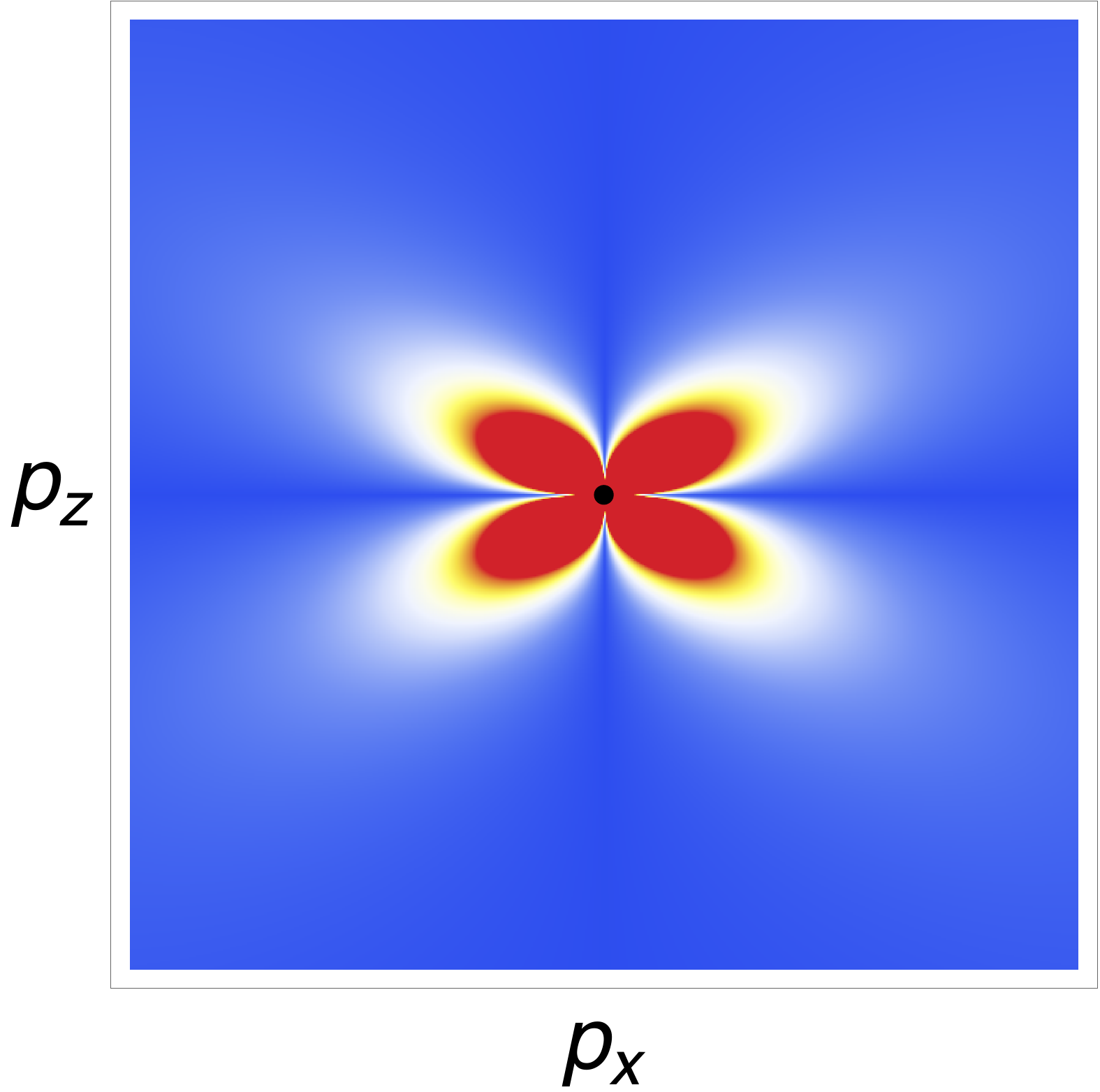}}\hspace{1cm}
\subfloat[]{\label{fig:berry-curvature-d-contour-plot-y-z}\includegraphics[scale=0.13]{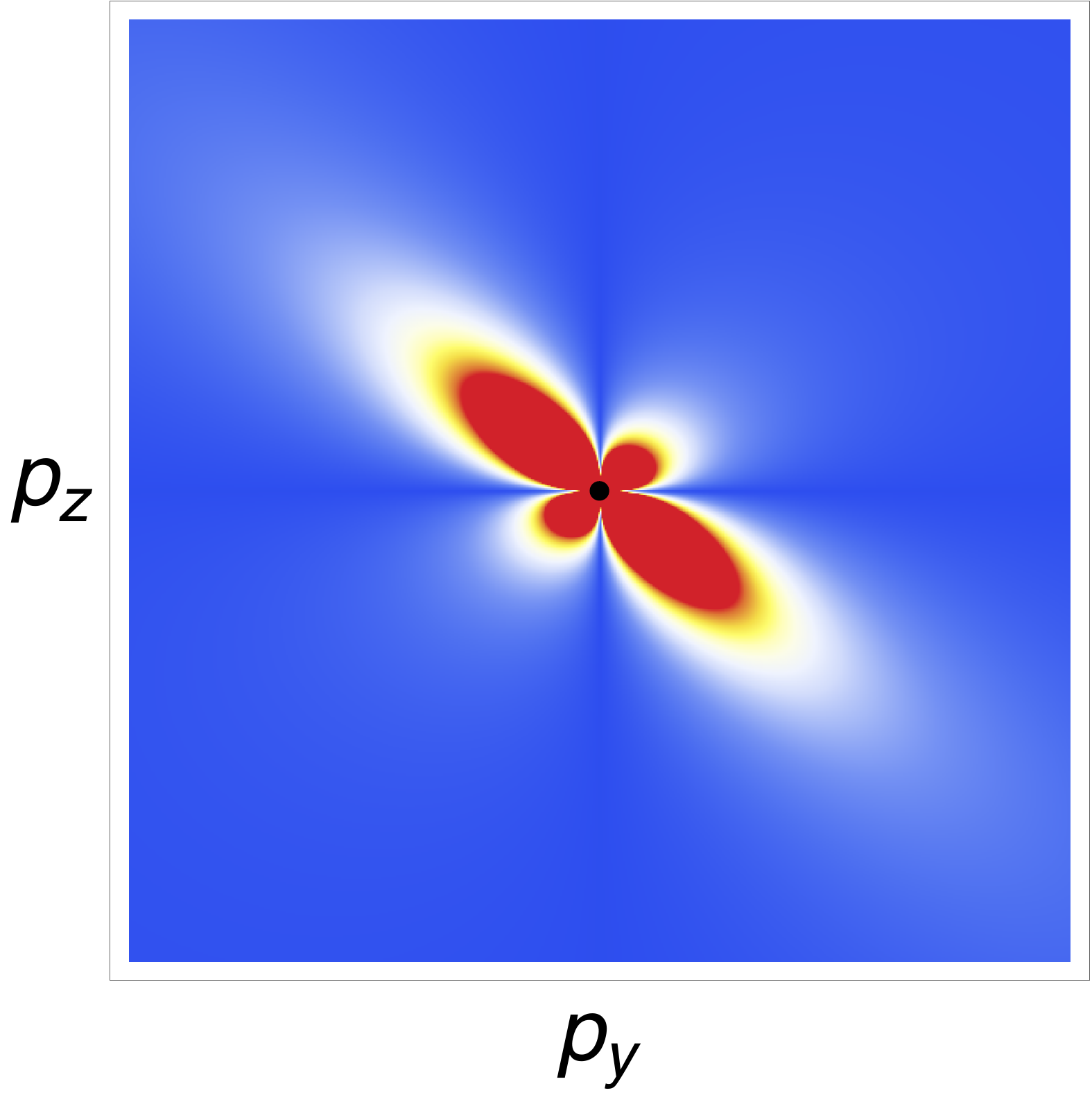}}
\caption{Contour plot of the 
curvature~\eqref{eq:berry-curvature-purely-spacelike-d}
\protect\subref{fig:berry-curvature-d-contour-plot-x-y} 
in the $p_x$-$p_y$ plane,
\protect\subref{fig:berry-curvature-d-contour-plot-x-z}
in the $p_x$-$p_z$ plane,
and \protect\subref{fig:berry-curvature-d-contour-plot-y-z}
in the $p_y$-$p_z$ plane.}
\label{eq:berry-curvature-d-contour-plots}
\end{figure*}

Finally, we explore the case of purely spacelike $d$ coefficients such that $d_{00}=d_{i0}=d_{0j}=0$ and $d_{ij}\neq 0$ for $j>i$. No additional time derivative appears, and the hamiltonian is 
\begin{equation}
H_d=H_0+d_{ij}\gamma^0\gamma_5\gamma^i\lambda_j\,,
\end{equation}
where $H_0$ is the Dirac hamiltonian of Eq.~\eqref{eq:dirac-hamiltonian-standard}. 

It is convenient to define the matrices $\mathds{D}^{(\pm)}$ with components $(\mathds{D}^{(\pm)})_{ij}:=\delta_{ij}\pm d_{ij}$. Then, the exact dispersions can be cast into the compact form
\begin{equation}
\label{eq:dispersions-d-spacelike}
E_u^{(1,2)}=|\mathds{D}^{(\mp)}\mathcal{p}|=E_v^{(1,2)}\,.
\end{equation}
The two Weyl cones are spatially distorted such that they exhibit several intersections with each other. See Figs.~\ref{fig:dispersions-d-spacelike} and \ref{fig:dispersions-d-spacelike-reinterpreted} for the case $d_{12}=d_{13}=d_{23}=3/4$. The $u$- and reinterpreted $v$-type spinor-valued energy eigenstates are
\begin{subequations}
\begin{align}
\label{eq:spinors-u-type-d}
|u^{(1,2)}(\mathcal{p})\rangle&=\mathcal{N}_u^{(1,2)}\begin{pmatrix}
\pm E_u^{(1,2)}+p_z \\
(\mathds{D}^{(\mp)}\mathcal{p})_x+\mathrm{i}(\mathds{D}^{(\mp)}\mathcal{p})_y \\
-E_u^{(1,2)}\mp p_z \\
\mp (\mathds{D}^{(\mp)}\mathcal{p})_x\mp \mathrm{i}(\mathds{D}^{(\mp)}\mathcal{p})_y \\
\end{pmatrix} \notag \\
&=|v^{(1,2)}(\mathcal{p})\rangle\,,
\end{align}
with the normalization factors
\begin{equation}
\mathcal{N}_u^{(1,2)}=\frac{1}{2\sqrt{E_u^{(1,2)}(E_u^{(1,2)}\pm p_z)}}\,.
\end{equation}
\end{subequations}

The Berry connection has an intricate form, but it can be expressed in a compact manner as follows:
\begin{subequations}
\label{eq:berry-connection-purely-spacelike}
\begin{equation}
\mathcal{A}_u^{(1,2)}=\frac{\mathcal{p}\times \zeta^{(1,2)}}{2E_u^{(1,2)}(E_u^{(1,2)}\pm p_z)}=\mathcal{A}_v^{(1,2)}\,,
\end{equation}
with the auxiliary vectors
\begin{equation}
\zeta^{(1,2)}:=\begin{pmatrix}
\pm d_{13}+d_{12}d_{23} \\
\pm d_{23} \\
1 \\
\end{pmatrix}\,.
\end{equation}
\end{subequations}
The Dirac strings run along sets where $E_u^{(1,2)}=\mp p_z$. These connections reduce to Eq.~\eqref{eq:berry-connection-u} for $d_{ij}=0$, modulo global signs and an interchange of the labels $(1,2)$. 

The curvatures take forms analogous to the standard ones of Eq.~\eqref{eq:berry-curvature-isotropic-b}, but with $p$ in the denominators replaced by the modified dispersion with the appropriate label,
\begin{equation}
\Omega_{u,i}^{(1,2)}=\mp\frac{p_i}{2(E_u^{(1,2)})^3}=\Omega_{v,i}^{(1,2)}\,.
\label{eq:berry-curvature-purely-spacelike-d}
\end{equation}
The signs appearing in Eq.~\eqref{eq:berry-curvature-purely-spacelike-d} and the interchange of labels can be understood by applying a parity transformation in momentum space, $\mathcal{p}\mapsto-\mathcal{p}$. This adapts the labeling and global signs of Eq.~\eqref{eq:berry-connection-purely-spacelike} for $d_{ij}=0$ to the ones of Eq.~\eqref{eq:berry-connection-u}. The sign change of the curvature under $\mathcal{p}\mapsto -\mathcal{p}$ is compensated by the additional sign change $\nabla_{\mathbf{p}}\mapsto -\nabla_{\mathbf{p}}$.

The sums of $\Omega_{u,i}^{(1,2)}$ and $\Omega_{v,i}^{(1,2)}$
produce nonzero results,
unlike the case of nonzero $d_{i0}$ discussed in Sec.~\ref{sec:mixed-d-1}.
To display these modifications, contour plots of the norms $|(\Omega_{u,i}^{(1)})+(\Omega_{u,i}^{(2)})|$ in three mutually orthogonal planes are presented in Fig.~\ref{eq:berry-curvature-d-contour-plots} for the nonzero values $d_{12}=d_{13}=d_{23}=3/4$ of the three off-diagonal spatial $d$-type coefficients. In two planes the curvature has a symmetrical butterfly shape, but in
the third the butterfly shape is distorted. These plots are reminiscent of quadrupole distributions. 

Due to the involved form of the dispersions, it is challenging to integrate the curvature analytically. However, numerical computations can be performed
to reveal that the first Chern numbers are 
\begin{equation}
N_u^{(1,2)}= N_v^{(1,2)}= \mp 1
\quad {\rm (spacelike~case)} \,.
\end{equation}

\subsection{Momentum-space geometric phases}

We can determine the momentum-space geometric phases for the various scenarios
with nonzero coefficients $d_{\mu\nu}$ along circles in $\mathcal{P}_3$. However, unlike the $b_{\mu}$ sector, the $d_{\mu\nu}$ coefficients do not separate the Weyl cones along a direction in $\mathcal{P}_4$. Instead, each Weyl cone is deformed differently, which also lifts the fourfold degeneracy. Projecting the circles in $\mathcal{P}_3$ onto these cones implies closed loops, but $E$ varies along these curves. 

For the isotropic case of Sec.~\ref{sec:isotropic-d} and the second mixed sector of Sec.~\ref{sec:mixed-d-2}, the curvatures match those of Eq.~\eqref{eq:berry-curvature-isotropic-b}. Integrating the latter around apices of the Weyl cones along circles in $\mathcal{P}_3$ and avoiding the Dirac strings according to the procedure of Sec.~\ref{sec:purely-timelike-sector} 
yields the momentum-space geometric phases
\begin{equation}
\Phi^{(1)}_u=-\pi=\Phi^{(1)}_v\,,\quad \Phi^{(2)}_u=\pi=\Phi^{(2)}_v\,.
\label{dphases}
\end{equation}
This result is in agreement with Eq.~\eqref{eq:berry-phases-timelike-case} even
though the loops on the Weyl cones in $\mathcal{P}_4$ are no longer circles of constant energy. The projections of these loops are shown in Figs.~\ref{fig:dispersions-d-isotropic-reinterpreted} and \ref{fig:dispersions-d-mixed-2-reinterpreted}. 

For the first mixed sector of Sec.~\ref{sec:mixed-d-1} and the purely spacelike regime of Sec.~\ref{sec:purely-spacelike-d},
the curvatures are modified by the Lorentz violation. The calculations of the integrals to obtain the momentum-space geometric phases therefore differ from those of the conventional Dirac theory. However, performing the computations explicitly reveals that the momentum-space geometric phases for these sectors are again given by Eq.~\eqref{dphases} modulo $2\pi$.

\section{Alternative invariants}
\label{sec:alternative-invariants}

Our treatment of topological properties 
above is based on the first Chern number.
This section presents some results on alternative topological invariants.

The gap nodes at the Fermi energy $E_F=0$, 
known as Fermi points,
can be characterized by the topological 
invariant~\cite{Klinkhamer:2004hg,Volovik:2003fe} 
\begin{subequations}
\label{eq:topological-invariant}
\begin{align}
N_a&=\mathrm{tr}\oint_{\Sigma_a}\mathrm{d}n^{\sigma}\,\mathcal{N}_{\sigma}\,, \displaybreak[0]\\[2ex]
\mathcal{N}_{\sigma}&=\frac{1}{24\pi^2}\varepsilon_{\mu\nu\varrho\sigma}S\frac{\partial\mathcal{D}}{\partial p_{\mu}}S\frac{\partial\mathcal{D}}{\partial p_{\nu}}S\frac{\partial\mathcal{D}}{\partial p_{\varrho}}\,.
\end{align}
\end{subequations}
Here,
$\mathrm{d}n^{\sigma}=\mathrm{d}\Sigma_a\,n^{\sigma}$ is the volume element of the closed three-dimensional hypersurface $\Sigma_a$ 
with unit 4-vector normal $n^{\sigma}$. Also, $\mathcal{D}=\mathcal{D}(\mathrm{i}\lambda_4,\lambda_k)$ is the Dirac operator in momentum space and $S=\mathcal{D}^{-1}(\mathrm{i}\lambda_4,\lambda_k)$ the Dirac propagator,
both being evaluated at imaginary energies $\lambda_0=\mathrm{i}\lambda_4$ via a Wick rotation. 
The trace is computed over spinor indices,
and $\varepsilon_{\mu\nu\varrho\sigma}$ is the totally antisymmetric Levi-Civita symbol.

The topological invariant $N_a$ takes integer values, $N_a\in\mathbb{Z}$, 
and is sensitive to isolated Fermi points. For $N_a=0$, the topology of the Fermi point is trivial. When $N_a\neq 0$, the topology of the bundle 
$P({\rm U(1)},\mathcal{P}_4)$
is nontrivial and the Fermi points are insensitive to small perturbations. 

Note that $N_a$ is evaluated in the energy-momentum space $\mathcal{P}_4$ with $(\lambda_0,\lambda_k)\in\mathcal{P}_4$. 
In contrast, both the Berry curvature and the momentum-space geometric phases
considered above are evaluated in $\mathcal{P}_3$. 
Also,
Eq.~\eqref{eq:topological-invariant} offers a computational advantage
in that it requires only the propagator in the Dirac sector of the SME. 
In particular,
knowledge of the eigenstates of the quantum system is unnecessary.

\subsection{CPT-odd case}

The Dirac propagator for each one of the coefficients of the SME fermion sector is known \cite{Reis:2016hzu,AndradedeSimoesdosReis:2019aim}. In particular, for the $b_{\mu}$ sector with a vanishing fermion mass, we cast the result into
\begin{subequations}
\label{eq:propagator-b}
\begin{align}
S_b(\lambda_0,\lambda_k)&=\frac{1}{\Delta}(\xi_{\mu}\gamma^{\mu}+\zeta_{\mu}\gamma_5\gamma^{\mu})\,, \displaybreak[0]\\[2ex]
\xi^{\mu}&=(\lambda^2+b^2)\lambda^{\mu}-2(b\cdot \lambda)b^{\mu}\,, \displaybreak[0]\\[2ex]
\zeta^{\mu}&=-2(b\cdot \lambda)\lambda^{\mu}+(\lambda^2+b^2)b^{\mu}\,, \displaybreak[0]\\[2ex]
\label{eq:denominator-propagator-b}
\Delta&=(\lambda-b)^2(\lambda+b)^2\,.
\end{align}
\end{subequations}
The trace of the matrix density $\mathcal{N}_{\sigma}n^{\sigma}$ of Eq.~\eqref{eq:topological-invariant} can be evaluated in covariant form,
\begin{align}
\mathrm{tr}(\mathcal{N}_{\sigma}n^{\sigma})&=\frac{1}{\pi^2\Delta^2}\tilde{\eta}_{\mu\nu}n^{\mu}\bigg\{4\lambda^{\nu}(b\cdot \lambda)(\lambda^2+b^2) \notag \\
&\phantom{{}={}}\hspace{0.8cm}-b^{\nu}\Big[4(b\cdot \lambda)^2+(\lambda^2+b^2)^2\Big]\bigg\}\bigg|_{\lambda_0\rightarrow \mathrm{i}\lambda_4}\,,
\end{align}
with $\tilde{\eta}_{\mu\nu}:=\mathrm{diag}(\mathrm{i},-1,-1,-1)$ and the Minkowski metric $\eta_{\mu\nu}$ is employed for all remaining scalar products.

First, let us evaluate $N_a$ for a purely timelike $b_{\mu}$. The hypersurface $\Sigma_a$ is chosen as a 3-sphere parametrized as
\begin{equation}
\begin{pmatrix}
\lambda_4 \\
\lambda_x \\
\lambda_y \\
\lambda_z \\
\end{pmatrix}=\lambda\begin{pmatrix}
\sin\theta\sin\phi\sin\chi \\
\sin\theta\sin\phi\cos\chi \\
\sin\theta\cos\phi \\
\cos\theta \\
\end{pmatrix}\,,
\end{equation}
with $\lambda:=\sqrt{\sum_i \lambda_i^2}$ and angles $\theta\in [0,\pi]$, $\phi\in [0,\pi]$, and $\chi\in [0,2\pi)$. Integrations around the Weyl nodes at $\lambda_4=\pm\mathrm{i}b_0$ provide $N_a=\mp 1$. To avoid singularities of the integrand, it is important to choose the radius of $\Sigma_a$ small enough such that the nodal circle of Fig.~\ref{fig:dispersions-timelike-b} is not contained in it. Similarly, we can consider a purely spacelike $b_{\mu}$ with $\mathcal{b}=(0,0,b_z)$, where integrations around the Weyl nodes $(\lambda_k)=\pm\mathcal{b}$ result in $N_a=\pm 1$, as well. Thus, we conclude that these points are topologically protected, which agrees with the results related to the curvatures of Eqs.~\eqref{eq:berry-connection-integrated}, \eqref{eq:berry-curvature-spacelike-case-1}, and \eqref{eq:berry-curvature-spacelike-case-2}. An integration over a 3-sphere around the origin $\lambda_k=0$ without enclosing one of the Weyl nodes leads to $N_a=0$, as expected.

The nodal circle of the isotropic $b_{\mu}$ sector,
which is visible in Fig.~\ref{fig:dispersions-timelike-b},
represents a Fermi surface that corresponds to a 2-sphere with one of the spatial momentum components disregarded. The topological properties of this contour must be determined differently from those of isolated Weyl or Dirac points. Here, integration must be performed along a closed curve $\gamma$ that encircles the nodal line. For a modified Dirac theory with Dirac operator $\mathcal{D}$ and propagator $S_F$, the corresponding topological invariant is~\cite{Volovik:2003fe}
\begin{equation}
\label{eq:topological-invariant-fermi-surface}
N_1=\mathrm{tr}\oint_{\gamma} \frac{\mathrm{d}\vartheta}{2\pi\mathrm{i}}S(\lambda_4,\lambda_k)\partial_{\vartheta}\mathcal{D}(\lambda_4,\lambda_k)\,,
\end{equation}
with the trace performed over the spinor indices. The curve $\gamma$ is chosen as a loop parametrized in terms of the angle $\vartheta\in [0,2\pi]$ and shifted appropriately such that it runs around the nodal circle. In particular, $\lambda_4=-\sin(\vartheta)$, where we choose $\lambda_x=0$, $\lambda_y=\cos\varphi(\cos\vartheta+b_0)$, and $\lambda_z=\sin\varphi(\cos\vartheta+b_0)$. Then, the nodal circle in the $\lambda_y$-$\lambda_z$ plane is enclosed by a loop centered at the point $\lambda_4=0$, $\lambda_y=b_0\cos\varphi$, and $\lambda_z=b_0\sin\varphi$ and orthogonal to the tangent vector of the nodal circle at this point.

Applying Eq.~\eqref{eq:topological-invariant-fermi-surface} with Eq.~\eqref{eq:propagator-b} inserted and $\lambda_0\mapsto \mathrm{i}\lambda_4$, we obtain that $N_1=0$. This holds because the propagator for the $b_{\mu}$ model, stated in Eq.~\eqref{eq:propagator-b}, involves the overall factor $\Delta^{-1}$, given by Eq.~\eqref{eq:denominator-propagator-b}. For the isotropic sector, $\Delta^{-1}$ is a product of scalar toy model propagators of the form
\begin{equation}
\label{fig:toy-model-propagators}
G_{\pm}(\lambda_4,\lambda_k)=\frac{1}{(\mathrm{i}\lambda_4\pm \lambda_F)^2-\lambda^2}\,,
\end{equation}
with $\lambda_0\mapsto \mathrm{i}\lambda_4$ and $\lambda_F$ interpreted as the Fermi momentum. The latter have Fermi surfaces at $\lambda_4=0$, $\lambda=\lambda_F$. Integration along a circle around $\lambda_x=\lambda_F$ implies $N_1^{\pm}=\mp 1$. Since the Fermi surfaces with $N_1^{\pm}=\mp 1$ are encircled simultaneously when the integral of Eq.~\eqref{eq:topological-invariant-fermi-surface} is evaluated for Eq.~\eqref{eq:propagator-b}, the nontrivial topological quantum numbers compensate each other: $N_++N_-=-1+1=0$. This means that the Fermi surface for the purely timelike sector of the $b_{\mu}$ model is topologically trivial, i.e., it is not protected against small perturbations. Indeed, introducing a small fermion mass opens a gap and the Fermi surface vanishes, confirming the computational outcome.

\subsection{CPT-even case}
\label{sec:alternative-invariants-d}

The propagator of the $d_{\mu\nu}$ sector has a form analogous to Eq.~\eqref{eq:propagator-b} for the $b_{\mu}$ coefficients \cite{Reis:2016hzu,AndradedeSimoesdosReis:2019aim}. However, the dependence on the four-momentum is more involved, 
\begin{subequations}
\label{eq:propagator-d}
\begin{align}
S_d(\lambda_0,\lambda_k)&=\frac{1}{\Delta}(\xi_{\mu}\gamma^{\mu}+\zeta_{\mu}\gamma_5\gamma^{\mu})\,, \displaybreak[0]\\[1ex]
\xi^{\mu}&=[\lambda^2+(d\cdot \lambda)^2]\lambda^{\mu}-2(\lambda\cdot d\cdot \lambda)d^{\mu\nu}\lambda_{\nu}\,, \displaybreak[0]\\[1ex]
\zeta^{\mu}&=2(\lambda\cdot d\cdot \lambda)\lambda^{\mu}-[\lambda^2+(d\cdot \lambda)^2]d^{\mu\nu}\lambda_{\nu}\,, \displaybreak[0]\\[1ex]
\label{eq:denominator-propagator-d}
\Delta&=(\lambda+d\cdot \lambda)^2(\lambda-d\cdot \lambda)^2\,,
\end{align}
\end{subequations}
where $d\cdot \lambda$ denotes $d^{\mu\nu}\lambda_{\nu}$. 

Using the above propagator,
we seek to evaluate the integral for the topological quantum number $N_a$ of Eq.~\eqref{eq:topological-invariant} for each of the cases discussed in Secs.~\ref{sec:isotropic-d} -- \ref{sec:purely-spacelike-d}.
The integrals are challenging to carry out for generic coefficients, so we perform computations for specific values. 
Note that the integration may be problematic when some SME coefficients become large.
For example, for $d_{i0}=-0.58$, the dispersion of Eq.~\eqref{eq:dispersion-relations-mixed-d-1} becomes complex in certain regions of momentum space, which renders the value of Eq.~\eqref{eq:topological-invariant} invalid as a topological charge. Similarly, for $d_{0i}=-0.58$ the dispersion of Eq.~\eqref{eq:dispersion-relations-mixed-d-2} is zero in particular regions, whereupon Eq.~\eqref{eq:topological-invariant} becomes singular. However,
no such issues occur for the purely spacelike coefficients $d_{ij}$ with $i\neq j$ because $\lambda_i(\mathds{D}^{(\pm)T})_{ik}(\mathds{D}^{(\pm)})_{kj}\lambda_j>0$ for $\lambda_k\neq 0$, which prevents complex or zero values of the dispersions of Eq.~\eqref{eq:dispersions-d-spacelike} from occurring beyond the Weyl nodes.

For the isotropic case, computation yields the results
\begin{equation}
N_a=\left\{\begin{array}{rc}
0 & 0\leq |d_{00}|<1 \,, \\[0.5ex]
\pm 1 & d_{00}=\mp 1 \,, \\[0.5ex]
\pm 2 & d_{00}\in (\mp 1,\mp 3) \,, \\[0.5ex]
\pm 1 & d_{00}=\mp 3 \,, \\[0.5ex]
0 & |d_{00}|>3\,.
\end{array}
\right.
\end{equation}
This indicates a trivial topology for $d_{00}$ lying in a finite, open interval centered around the Lorentz-invariant case. 
At $|d_{00}|=1$ the dispersion~\eqref{eq:dispersion-relations-isotropic-d}
becomes singular,
which is reflected in the apparent topology change.
Note also that the form~\eqref{eq:inverse-A-isotropic-d} of the matrix $A$, which was introduced to remove the additional time derivative from the Lagrange density appearing in this sector, is inapplicable for $|d_{00}|>1$. 
Another topology change is seen to occur at $|d_{00}|= 3$,
where one of the two dispersions of Eq.~\eqref{eq:dispersion-relations-isotropic-d} vanishes. The corresponding fermion mode thus cannot provide any contribution to $N_a$, which reduces it by one unit to $|N_a|=1$. 

For the mixed coefficients $d_{i0}$, $d_{0i}$ and the purely spacelike ones $d_{ij}$, we obtain $N_a=0$ for values within a range around zero. Each of these sectors can thus be interpreted as topologically trivial. This result follows because the Weyl nodes are joined, and the opposite topological charges of the individual nodes therefore cancel each other.

\section{Summary and outlook}
\label{sec:conclusions}

This work concerns the effect of a background
on the momentum-space geometric phases
associated with the relativistic motion 
of massless spin-$\tfrac 12$ Weyl fermions in Lorentz-violating models, 
where the role of the background
is played by the coefficients for Lorentz violation.
Two examples are studied.
Section~\ref{sec:application-sme-b}
discusses the scenario with CPT violation involving a coefficient $b_\mu$,
while Sec.~\ref{sec:application-sme-d}
considers the CPT-preserving case with a coefficient $d_{\mu\nu}$.
The topological properties of the momentum-space geometric phases
are investigated using the Berry curvature and the first Chern number.
Alternative topological invariants are explored
in Sec.~\ref{sec:alternative-invariants}.

When a nonzero coefficient $b_\mu$ is present,
our results demonstrate that 
significant changes occur to the Berry curvature when the two Weyl cones become nondegenerate along a spatial direction in momentum space. Then, the source and sink of the curvature, which coincide in the Lorentz-invariant regime, are separated and a Berry flux emerges in the plane orthogonal to the line of separation. Momentum-space geometric phases of $\pm\pi$ can accumulate when the apices of the Weyl cones are encircled during the fermion motion. These geometric phases play a special role for occupied fermion and antifermion states at $E=0$, which are of particular relevance
in certain scenarios with large Lorentz violation.

In the presence of a nonzero $d_{\mu\nu}$ coefficient,
the Weyl cones remain joined in momentum space, but they can be subject to tilts or spatial distortions. Our results demonstrate that a tilt leaves unaffected the Berry curvature, whereas spatial distortions of Weyl cones are reflected in curvature distortions. The norm of the total curvature for $u$- and $v$-type spinors no longer vanishes, and lines of constant curvature acquire a butterfly shape. The topological charges of the Weyl nodes remain $\pm 1$.
Momentum-space geometric phases of $\pm\pi$ emerge along loops encircling the apices of Weyl cones.

Our investigations reveal that the geometry of the {\rm U(1)} bundle of energy eigenstates over momentum space remains unaffected when the Weyl cones are shifted along the energy axis, when they are tilted, or when their opening angle changes. However, a separation of the Weyl cones along a spatial-momentum direction and spatial deformations both impact the geometry. 
These results are consistent with our studies of certain topological invariants
other than the first Chern number,
which confirm that topological charges are present at degenerate points in energy-momentum space. These invariants are determined based on the propagator of a theory, rather than the hamiltonian and its eigenstates.

The results in this work represent an initial step 
toward the establishment of a comprehensive treatment
of momentum-space geometric phases in the full SME.
Several direct theoretical extensions of our analysis can be envisaged
for future work.
An immediate possibility of definite interest 
is the extension of the results to other coefficients in the fermion sector,
including nonminimal ones.
Applications of these ideas to other sectors of the SME
are also of prospective interest.

Another area for exploration is associated with lifting the assumption
of time-independent and spatially homogeneous coefficients.
In practical applications to fundamental physics,
the cartesian coefficients for Lorentz violation
are typically assumed to be constant in a canonical 
Sun-centered frame~\cite{Bluhm:2003un,Bluhm:2001rw,Kostelecky:2002hh},
which provides an approximately inertial reference frame
over applicable experimental timescales.
However,
experimental measurements are performed in a noninertial frame
on or near the Earth's surface,
which implies time-dependent variations
in the observable coefficients for Lorentz violation~\cite{Kostelecky:1997mh}.
These variations oscillate at harmonics of the Earth's
sidereal frequency
and hence are adiabatic on typical experimental timescales.
A theoretical treatment of the implications of these oscillations
for observable geometric phases would be of definite interest.

Applications in the context of condensed-matter physics also have worthwhile prospects.  Consider, for example, a $b_\mu$-type Weyl semimetal of finite volume, in which one or more spatial components of $b_\mu$ are nonzero within a certain spatial domain $D$ but vanish outside $D$.  On the boundary $\partial D$, the topology of the ground state changes from nontrivial to trivial, implying the existence of interesting surface states such as Fermi arcs or drumhead surface states. A nontrivial ground-state topology may also modify measurable transport properties.

Other novel experimental applications of our work may also be feasible.
In the presence of Lorentz violation,
the value of the momentum-space geometric phase
is fixed for each nondegenerate Weyl cone.
Also, when the Weyl cones are separated or distorted 
along spatial-momentum directions, 
the total Berry curvature differs from the zero value 
of the Lorentz-invariant case.
These properties therefore establish the momentum-space geometric phase 
and the Berry curvature as gauge-invariant observables for Lorentz violation.
The experimental search for spin-nondegenerate Lorentz violation
governed by $b_{\mu}$ and $d_{\mu\nu}$ can thus be reinterpreted
as the search for fixed values of the momentum-space geometric phases
of massless spin-$\tfrac12$ fermions.
In particular,
even infinitesimal background coefficients suffice
to generate a nontrivial effect,
so the detection of Lorentz violation via momentum-space geometric phases
is unhampered by experimental sensitivity to the size of the coefficients.
Adapting this result to implementing a realistic experimental scenario
for a magnitude-independent detection of Lorentz violation
is an open challenge of considerable prospective interest.

\section*{Acknowledgments}

This work is supported in part 
by the U.S.\ Department of Energy 
under grant {DE}-SC0010120,
by grants FAPEMA Universal 00830/19 
and CNPq Produtividade 310076/2021-8,
by CAPES/Finance Code 001,
and by the Indiana University Center for Spacetime Symmetries.

\bibliography{paper-arXiv}

\end{document}